\documentclass[12pt]{iopart}

\usepackage{iopams}  
\usepackage{color}
\usepackage{graphicx}
\begin{document}

\title{Transition of blue-core helicon discharge}
\author{L. Chang$^{1\ast}$, S. J. Zhang$^1$, J. T. Wu$^2$, Y. W. Zhang$^3$, C. Wang$^1$, Y. Peng$^2$, S. S. Gao$^2$, C. J. Sun$^2$, Q. Wang$^2$, C. F. Sang$^{2\sharp}$, S. C. Thakur$^4$, S. Isayama$^5$, S. J. You$^6$}
\address{$^1$School of Electrical Engineering, Chongqing University, Chongqing, 400044, China}
\address{$^2$Key Laboratory of Materials Modification by Laser, Ion and Electron Beams (Ministry of Education), School of Physics, Dalian University of Technology, Dalian 116024, China}
\address{$^3$Institute of Plasma Physics, Hefei Institutes of Physical Science, Chinese Academy of Sciences, Hefei 23003, China}
\address{$^4$Department of Physics, Auburn University, Auburn, AL36849, United States of America}
\address{$^5$Department of Advanced Environmental Science and Engineering, Kyushu University, 6-1 Kasuga-Kohen, Kasuga, Fukuoka 816-8580, Japan}
\address{$^6$Applied Physics lab for PLasma Engineering (APPLE), Department of Physics, Chungnam National University, Daejeon 34134, Republic of Korea}
\ead{leichang@cqu.edu.cn, sang@dlut.edu.cn}

\date{\today}

\begin{abstract}
This study explores the transitional characteristics of blue-core helicon discharge, which to our knowledge was not particularly focused on before. Parameters are measured on recently built advanced linear plasma device, i.e. Multiple Plasma Simulation Linear Device (MPS-LD, Sun \textit{et al} 2021 \textit{Fusion Eng. Des.} \textbf{162} 112074 and Wu \textit{et al} 2024 \textit{Plasma Sources Sci. Technol.} \textbf{33} 085007) by various diagnostics including Langmuir probe, optical emission spectrometer, and standard high-speed camera. It is found that the jump direction of electron density (from low level to high level) is opposite to that of electron temperature (from high level to low level). Electron density increases significantly and the radial profile becomes localized near the axis when the blue-core transition occurs. With increased field strength, electron density increases whereas electron temperature drops. The radial profile of electron temperature looks like a ``W" shape, i.e. minimizing around the edge of blue-core column. Electron density increases with background pressure, while electron temperature peaks around certain pressure value. High-speed videos show that the plasma column oscillates radially and experiences azimuthal instabilities with high rate once entered blue-core mode. An electromagnetic solver (EMS) based on Maxwell's equations and a cold-plasma dielectric tensor is also employed to compute the wave field and power absorption during blue-core transition, to provide more details that are valuable for understanding the transitional physics but not yet available in experiment. The results show that wave field in both radial and axial directions changes significantly during the transition, its structure differs from antenna to downstream, and the power dependence of wave magnetic field is overall opposite to that of wave electric field. This work presents comprehensive characteristics of the transitional blue-core discharge and is important to both physics understanding and practical applications. 

\end{abstract}


\maketitle

\section{Introduction}
Helicon discharge is a low-temperature radio frequency (RF) plasma source that has been applied to semiconductor etching, material cleaning, surface treatment, space propulsion, heating and current drive for magnetic confinement fusion, and fundamental plasma physics research, etc\cite{Boswell:1970aa, Boswell:1997aa, Chen:1997aa, Yamada:2008aa, Chen:2015aa, Shinohara:2018aa, Takahashi:2019aa, Shinohara2022, Chang:2022aa, Chang2024}. The special features of helicon discharge include high plasma density (up to $10^{20}~\textrm{m}^{-3}$), high ionization ratio ($100\%$ in the core region), no plasma-contacting electrode, and low requirement for magnetic field (compared with electron cyclotron resonance (ECR) plasma). Blue-core mode is the highest level of helicon discharge using argon gas, i.e. the most popular helicon plasma source, and has been attracting great attention recently\cite{Zhang:2022aa, Lu:2022aa, Xia:2023aa, Wang:2023aa, Zhang:2024aa, Zhang:2024ab, Cui:2024aa, Stevenson:2024aa}. Here, the phrase of "highest level of helicon discharge" means that due to ion pumping effects higher power only yields higher ionisation states (like in a tokamak) rather than further density jump\cite{Chang:2022aa}.  Its underlying physics remain one of the most challenging topic for helicon community\cite{Chang:2022aa, Chang2024}, and has great potential for the thermal load test of plasma-facing material of fusion reactors (e.g. MPEX, material plasma exposure experiment\cite{Rapp:2016aa, Rapp:2020aa}) and high-power space electric thruster (e.g. VASIMR, variable specific impulse magnetoplasma rocket\cite{Chang-Diaz:2000aa}). Therefore, it is of both scientific value and practical significance to study the formation procedure of blue-core helicon discharge, especially the transitional characteristics and driven mechanisms from non-blue-core mode to blue-core mode, which are the focus of present work. Our motivation is to reveal the physics and guide the optimization of blue-core helicon discharge through measuring in detail the transitional features, analyzing its parameter dependence, and numerical computations based on a well benchmarked electromagnetic solver (EMS)\cite{Chen:2006aa}. Important findings include: the jump direction of electron density is opposite to that of electron temperature; the radial profile of electron temperature looks like a ``W" shape, i.e. minimizing around the edge of blue-core column; plasma column oscillates radially and experiences azimuthal instabilities with high rate once entered the blue-core mode; wave field changes significantly during the transition, its structure differs from antenna to downstream, and the power dependence of wave magnetic field is overall opposite to that of wave electric field. These are valuable to understanding the physical mechanism of blue-core helicon discharge and its practical applications. To our knowledge, it is the first research particularly focusing on the transition of blue-core helicon discharge. 

\section{Experimental Setup}
\subsection{Platform}
\begin{figure*}[ht]
\begin{center}
\includegraphics[width=1\textwidth,angle=0]{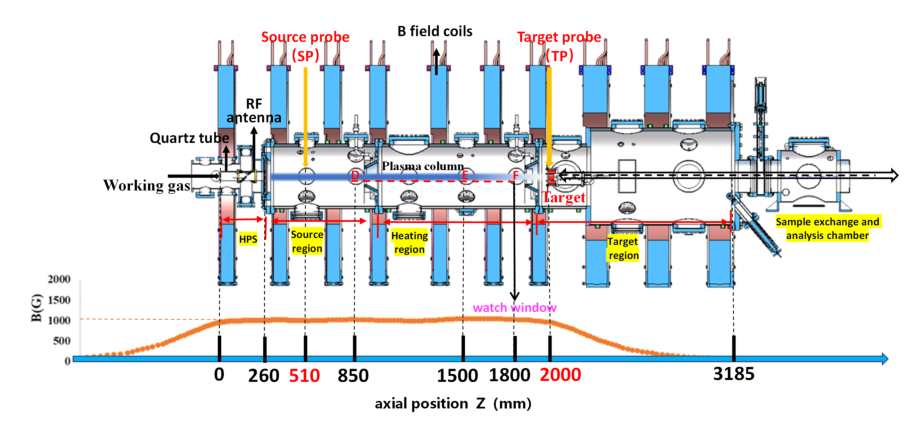}
\end{center}
\caption{A schematic of Multiple Plasma Simulation Linear Device (MPS-LD) (with permission from Plasma Sources Sci. Technol. 33 (2024) 085007).}
\label{fg_mps-ld}
\end{figure*}

To explore the underlying physics of blue-core helicon discharge and make our findings generally applicable, we choose a typical helicon device that was constructed recently and well tested, i.e. the Multiple Plasma Simulation Linear Device (MPS-LD)\cite{Sun:2021aa, Wu:2024aa}. It is designed mainly for the study of fusion-relevant plasma material interaction and diverter physics, and is an advanced platform to investigate the fundamental physics of helicon discharge, because of highly repeatable discharge and comprehensive diagnostics. Figure~\ref{fg_mps-ld} shows a schematic.
The device has three main components: vacuum system, magnet system and plasma source. The vacuum system includes source region, heating region, target region and sample exchange chamber. The magnet system has $11$ solenoid coils with independent power supplies, providing high flexibility of magnetic field configuration. The plasma source is fed by helicon discharge which is driven by a water-cooled copper half-turn helical antenna and $13.56$~MHz RF power supply. The formed plasma is first contained by a quartz tube with diameter of $5.5$~cm and length of $26$~cm, and then diffuses into the source and heating chamber with inner diameter of $40$~cm and the target chamber with inner diameter of $60$~cm. The total length of the plasma plume can be up to $3.185$~m, before the sample exchange chamber. The super long size makes this device particularly suitable for studying the axially decaying physics of helicon discharge, e.g. the limit of blue-core length. The coordinate system is also labelled in Fig.~\ref{fg_mps-ld}, with $z=0$~m the left end of the quartz tube and $z=3.185$~m the right end of the target chamber. The employed confining magnetic field is incorporated in Fig.~\ref{fg_mps-ld} too, which is overall uniform for helicon discharge and plasma diffusion regions. 

\subsection{Diagnostics}
In the present research, we make use of $3$ diagnostic systems to measure in detail the transitional features of blue-core helicon discharge, i.e. RF compensated Langmuir probe (LP) for electron density and temperature, optical emission spectrometer (OES) for particle species and their density and temperature, and standard high-speed camera for spatiotemporal discharge evolutions. Specifically, the Langmuir probe is a triple-probe system and gives data based on multiple collection averages under the same experimental parameters and with errors below $10\%$. It is located at $z=0.51$~m in the source region to measure the radial profiles of electron density and temperature that close to the helicon discharge. The OES is also located at $z=0.51$~m, through the opposite window of Langmuir probe, and is an Omni-$\lambda$750i-series grating monochromator or spectrograph. It has dual outlets that can be configured with two charge-coupled devices at the same time, and its side entrance can be connected with electronic shutter. It is designed with C-T structure and toroidal image calibration. The employed multi-grating tower design, which covers the full band of UV-VIS-IR spectral range, is very flexible for choosing the spectral range and resolution as need. The grating adopts $68\times 68$~mm large area grating, which improves the light collection efficiency significantly. The stray light rejection ratio is up to $10^{-5}$. Digital signal controllers (DSC) chip control is also utilized to make the choice of multiple entrances and exits more flexible, and double entry and exit can be selected according to needs. To enhance the accuracy of positioning, double entry and exit control through computer automatic control software are incorporated. Detailed parameters are listed in Tab.~\ref{table_oes}. 
\begin{table}[ht]
\caption{\label{table_oes} Key parameters of the employed optical emission spectrometer.}
\footnotesize
\begin{tabular}{ll}
\br
Name&Value\\
\mr
Focal Length (mm)&$750$\\
Relative Aperture&F/9.7\\
Optical Structure&C-T\\
Sweep Step (nm)&$0.005$\\
Stray Light&$1\times 10^{-5}$\\
Focal Plane&$30 (\textrm{w})\times 14 (\textrm{h})$\\
Height of Optical Axis (mm)&$146$\\
Grating Specification (mm)&$68\times 68$\\
Grating Station&Triple Grating\\
Slit Specification&Slit Width of $0.01-3$~mm With Continuous \& Manual Adjustable\\
Dimensions (mm)&$800\times 338\times 230$\\
Weight (kg)&$32.5$\\
Power Consumption&Maximum $100 \textrm{W}@24 \textrm{V}$\\
Communication Interface&USB2.0/RS-232\\
\br
\end{tabular}\\
\end{table}
\normalsize
The standard high-speed camera is Phantom-Series VEO $1310$, facing the observation window at $z=0.85$~m. It has resolution of $1280\times 960$ for $10860$ frames per second. The minimum exposure time is $1\mu$~s. The ultra-high sensitivity (ISO-12232 SAT) is $1.25\times 10^5$ for black and white image and $3.2\times 10^4$ for colored image. It comes with eraserable memory card (CFast 2.0 Type S) with maximum storage of $512$~GB. The minimum time interval between two exposures is $726$~ns. Image-based auto trigger (IBAT) is triggered automatically when the image changes. The fan can be also switched off in silent mode to eliminate noise and vibration. Detailed parameters are listed in Tab.~\ref{table_iccd}. Retarding potential analyzer (RPA) was used as well to measure the energy distribution functions of ions and electrons. Due to the low scanning voltage employed and strong interfering noise encountered (from RF power supply), however, the RPA data are not reliable and thereby not presented here. In our future work, we shall present the energetic particles physics of blue-core helicon discharge using the upgraded RPA diagnostics through solving these problems. 
\begin{table}[ht]
\caption{\label{table_iccd} Key parameters of the employed standard high-speed camera.}
\footnotesize
\begin{tabular}{ll}
\br
Name&Value\\
\mr
Maximum Resolution&$1280×\times 960@10860$~fps\\
Pixel Number&$1.2288\times 10^6$\\
Maximum Speed (fps)&$4.2335\times 10^5@320\times 24$\\
Continuously Adjustable Resolution&$640\times 12$\\
Minimum Exposure Time ($\mu$s)&$1$\\
PIV Specification& Time Interval $726$~ns $\&$ With The Support of Burst Mode\\
Sensor Type& CMOS (with correlated double sampling)\\
Sensor Size (mm)&$23\times 17.2$\\
Pixel Size ($\mu$m)&18\\
Sensitivity&$1.25\times 10^5$(black and whitte)/$3.2\times 10^4$(color)\\
Software/System&Phantom$\circledR$Camera Control/Windows\\
Power Supply&$100-240$~VAC\\
Power Consumption (W)& 65\\
Weight (kg)&2.3\\
Volume (cm)&$12.7\times 12.7\times 12.7$\\
\br
\end{tabular}\\
\end{table}
\normalsize

\subsection{Parameters}
In our experiment, argon gas is fed from left end and pumped towards the right, through the whole MPS-LD machine, with flow rate ranging from $40$ to $100$~sccm. The corresponding background pressure is from $0.13$~Pa to $0.29$~Pa. The strength of confining magnetic field is between $400$~G and $1500$~G, covering the typical magnitudes required for blue-core helicon discharge. The input RF power is also limited to the range of $100-2500$~W to cover typical blue-core helicon discharge parameters. 

\section{Experimental Results}
Based on the experimental platform, diagnostic devices and conditions presented above, we measured the transitional features of helicon discharge from non-blue-core mode to blue-core mode in great detail. Our strategy is to vary one control parameter among the input RF power, external magnetic field and flow rate while keep others fixed for one round of experiments, and observe the spatiotemporal evolutions of plasma in terms of electron density, electron temperature, emission spectrum, and optical image. All parameters are chosen in a narrow range around the critical value for blue-core mode transition, to focus on this particular stage during helicon discharge. 

\subsection{Effects of Input Power}
\begin{figure*}[ht]
\begin{center}$
\begin{array}{ll}
(a)&(b)\\
\includegraphics[width=0.493\textwidth,angle=0]{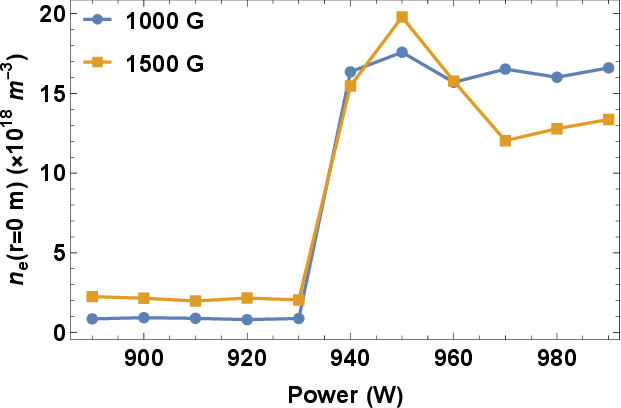}&\includegraphics[width=0.48\textwidth,angle=0]{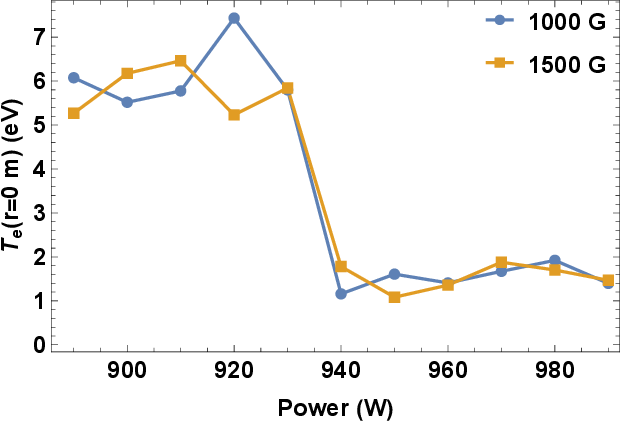}\\
\end{array}$
\end{center}
\caption{Transition features of blue-core helicon discharge for two external magnetic field strengths in terms of (on-axis): (a) electron density, (b) electron temperature.}
\label{fg_trans_com}
\end{figure*}

\begin{figure*}[ht]
\begin{center}$
\begin{array}{l}
(a)\\
\includegraphics[width=0.9\textwidth,angle=0]{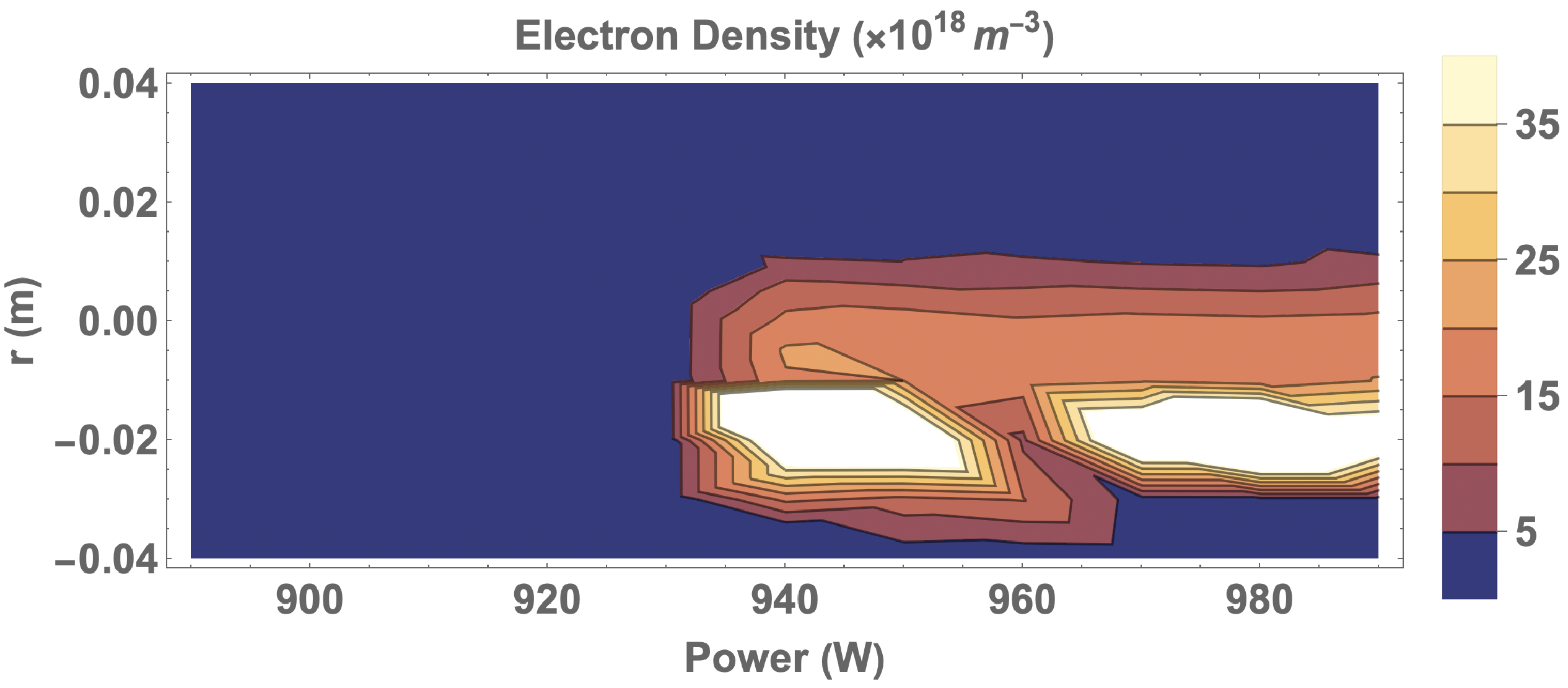}\\
(b)\\
\includegraphics[width=0.9\textwidth,angle=0]{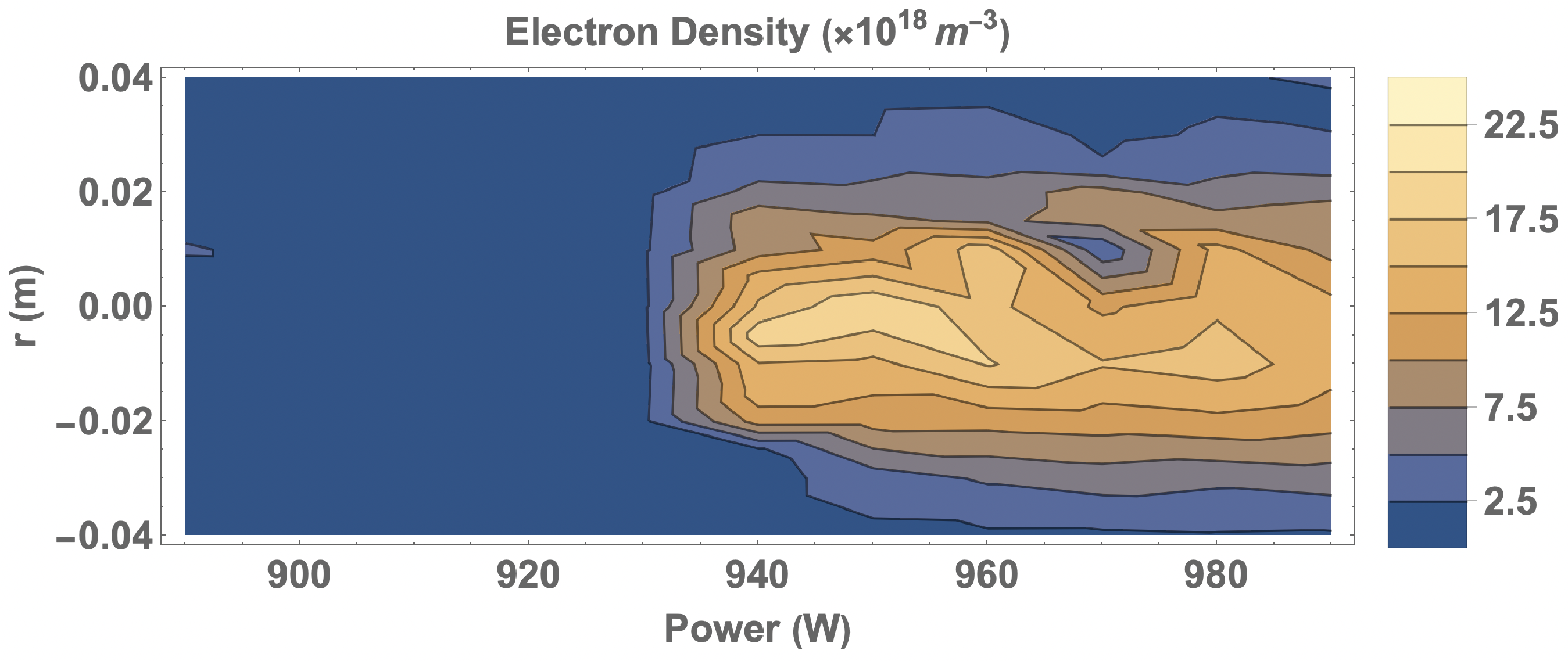}
\end{array}$
\end{center}
\caption{2D evolutions of the radial profiles of electron density for different input RF power magnitudes and external magnetic field strengths: (a) 1000 G, (b) 1500 G.}
\label{fg_map_ne}
\end{figure*}

\begin{figure*}[ht]
\begin{center}$
\begin{array}{l}
(a)\\
\includegraphics[width=0.9\textwidth,angle=0]{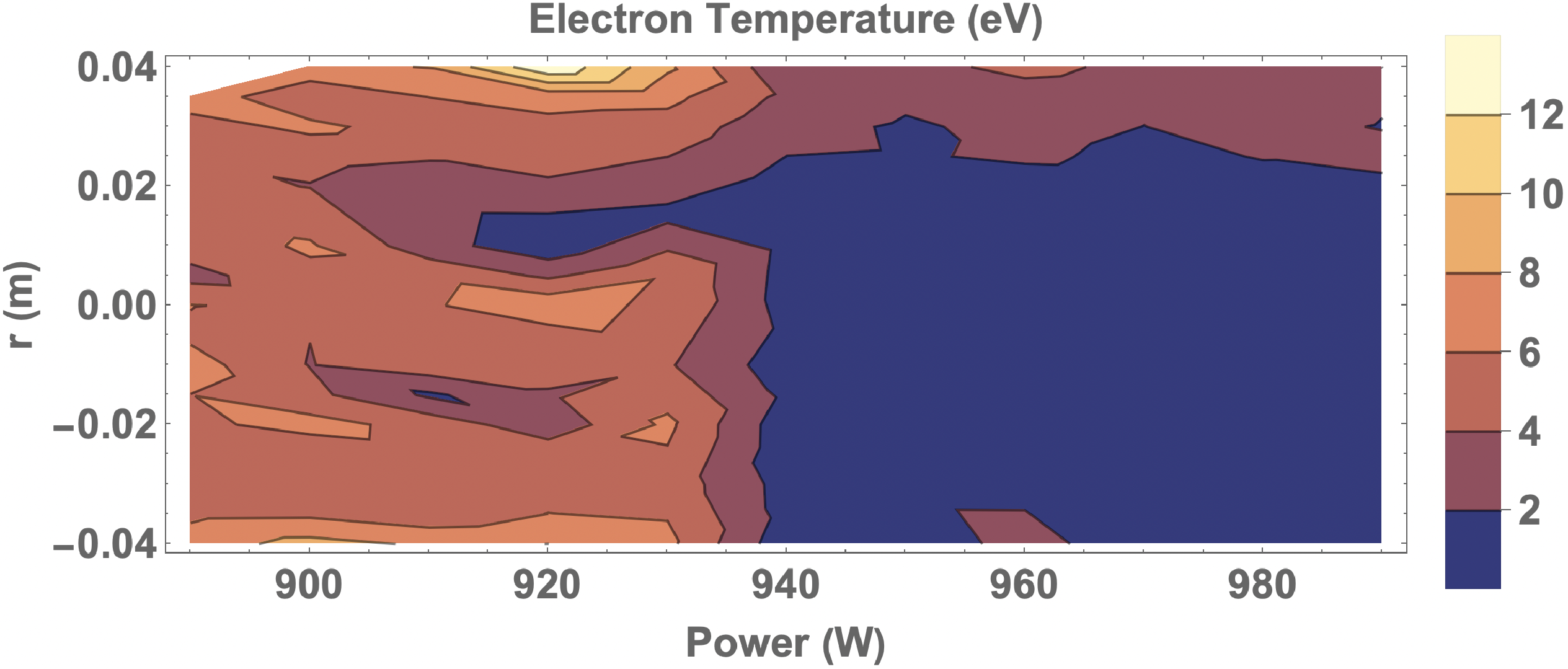}\\
(b)\\
\includegraphics[width=0.9\textwidth,angle=0]{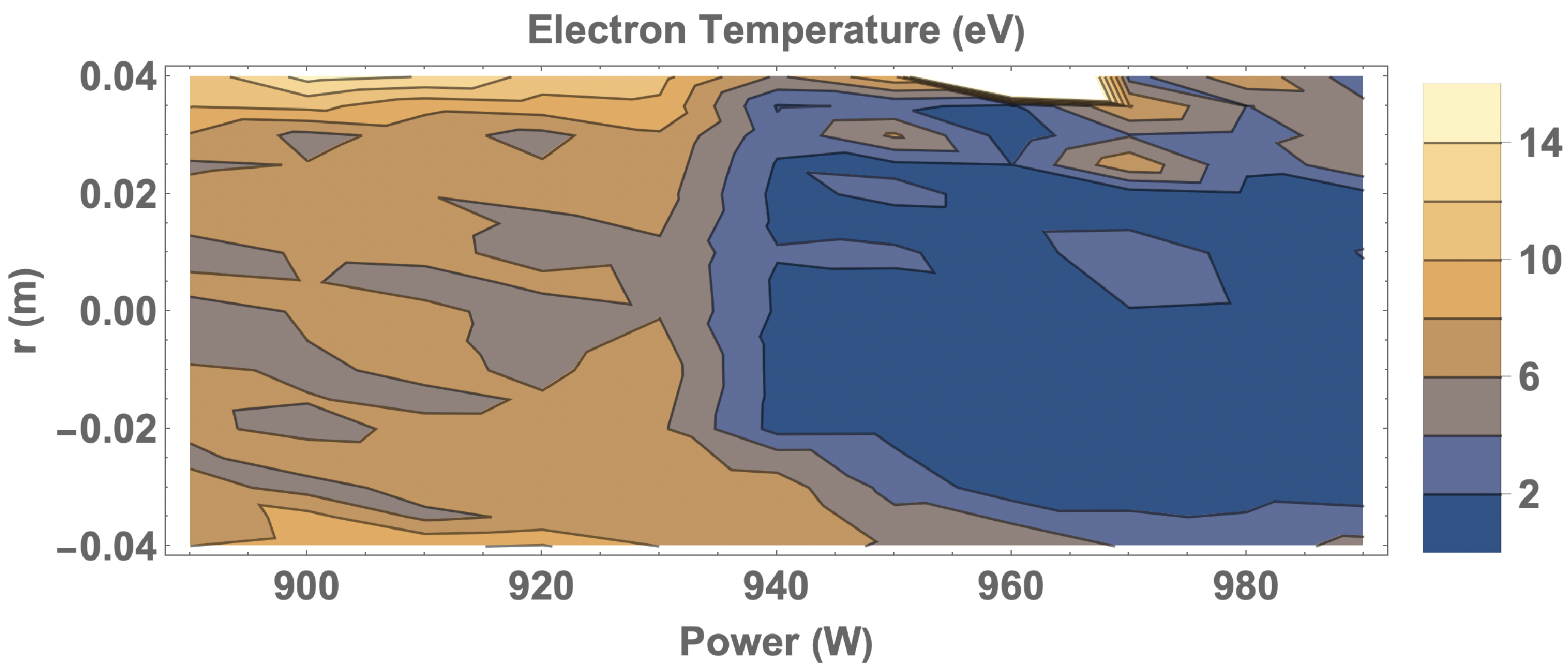}
\end{array}$
\end{center}
\caption{2D evolutions of the radial profiles of electron temperature for different input RF power magnitudes and magnetic field strengths: (a) 1000 G, (b) 1500 G.}
\label{fg_map_te}
\end{figure*}

We first explore the blue-core mode transition by varying the input RF power, i.e. the most convenient parameter, for two typical field strengths of $1000$~G and $1500$~G, respectively. The flow rate is fixed to $60$~sccm, maintaining background pressure of $0.19$~Pa. Figure~\ref{fg_trans_com} shows the measured results in terms of electron density and electron temperature (on axis).
It can be seen that the transition occurs around $930$~W for both of these two field strengths. Moreover, the jump direction of electron density (from low level to high level) is opposite to that of electron temperature (from high level to low level). The jumped ratios (ratio of magnitude before and after blue-core transition) for electron density is about $19.25$ while it is $0.23$ for electron temperature. This opposite trend has been observed in many previous studies\cite{Clarenbach2007, Chen2007a, Chen2012, Hu2020, Yuan:2020aa, Chang2021}, and it could be attributed to the competition between ionization procedure to vary plasma density and heating procedure to change plasma temperature. Indeed, if the excited and ionized argons are produced via electron impact collisions, which is applicable for helicon discharge\cite{Boswell:1970aa, Boswell:1997aa, Chen:1997aa}, the electron density is significantly affected by the electron temperature. Noticeably, higher magnetic field yields higher electron density before the blue-core mode transition, which is consistent with a previous study claiming that higher field requires higher jump threshold density according to Eq. (13)-(14) in \cite{Wu:2022aa}, whereas this relative magnitude is reversed after the transition. The magnitude of electron temperature does not change much when the field strength increases from $1000$~G to $1500$~G and this applies to both the stages of before and after the blue-core transition. Full pictures including the radial evolution of this blue-core plasma transition are given in Fig.~\ref{fg_map_ne} and Fig.~\ref{fg_map_te} for electron density and electron temperature, respectively. We can see that the density magnitude increases significantly and the radial profile becomes more shrunk and localized near the axis when the blue-core transition occurs. Interestingly, the peak density appears slightly off axis rather than on the axis. This (together with Fig.~\ref{fg_mf}(a) and Fig.~\ref{fg_trans_fl}(a)) implies radial asymmetry of helicon discharge on MPS-LD. Differently, the temperature magnitude decreases sharply from non-blue-core mode to blue-core mode, while the radial profile remains largely unchanged (this could be seen from the 2D map in Fig.~\ref{fg_map_te}(b) or more easily from the curves in Fig.~\ref{fg_mf}(b)). The overall ratio of measured results for two different field strengths indicated by these 2D maps is the same to that from the on-axis curves (Fig.~\ref{fg_trans_com}). 

\begin{figure*}[ht]
\begin{center}$
\begin{array}{ll}
(a_1)&(b_1)\\
\includegraphics[width=0.5\textwidth,angle=0]{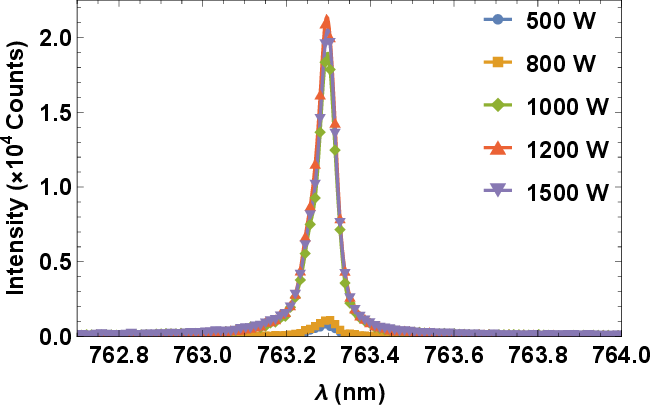}&\includegraphics[width=0.465\textwidth,angle=0]{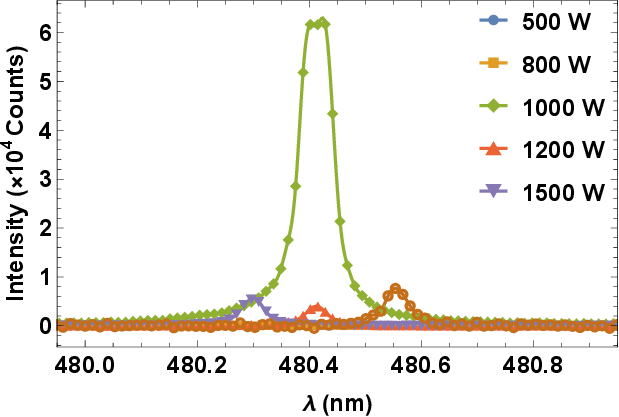}\\
(a_2)&(b_2)\\
\includegraphics[width=0.5\textwidth,angle=0]{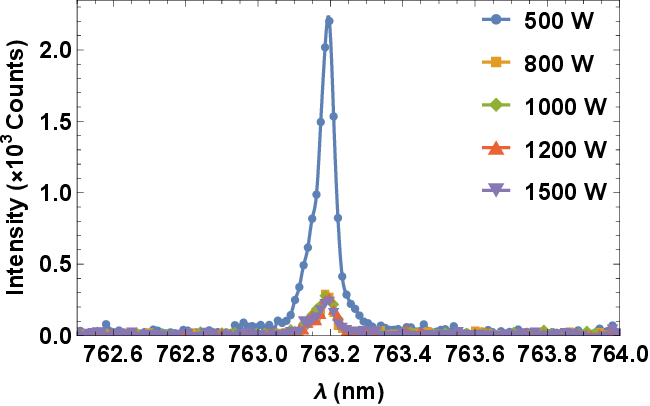}&\includegraphics[width=0.465\textwidth,angle=0]{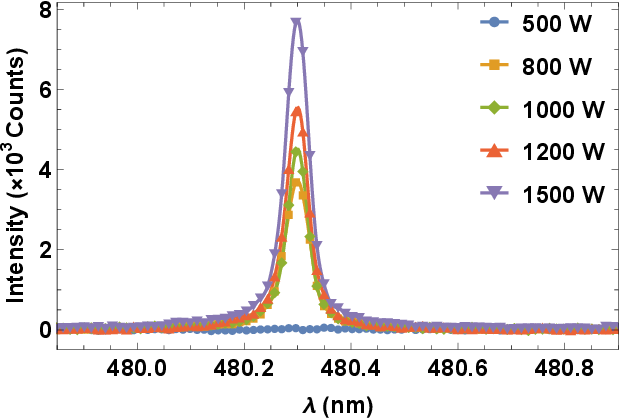}\\
(a_3)&(b_3)\\
\includegraphics[width=0.5\textwidth,angle=0]{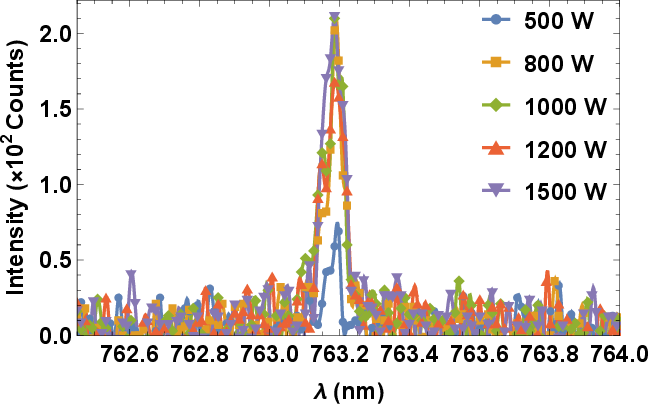}&\includegraphics[width=0.465\textwidth,angle=0]{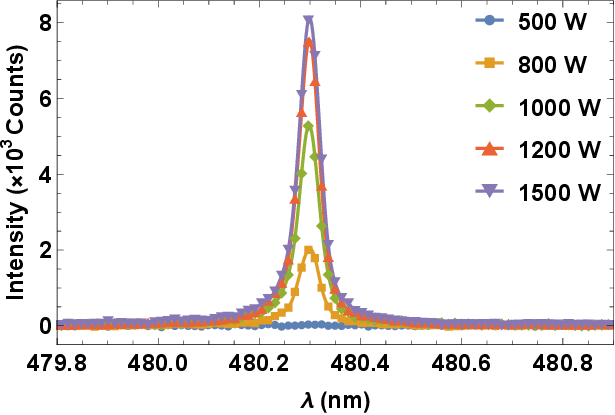}\\
\end{array}$
\end{center}
\caption{Measured optical emission spectrometer (OES) intensities of the ArI (left column) and ArII (right column) for different power levels ($500-1500$~W) and magnetic field strengths (top row: $600$~G, middle row: $800$~G, bottom row: $1000$~G).}
\label{fg_oes_power}
\end{figure*}

\begin{figure*}[ht]
\begin{center}$
\begin{array}{l}
(a)\\
\includegraphics[width=0.5\textwidth,angle=0]{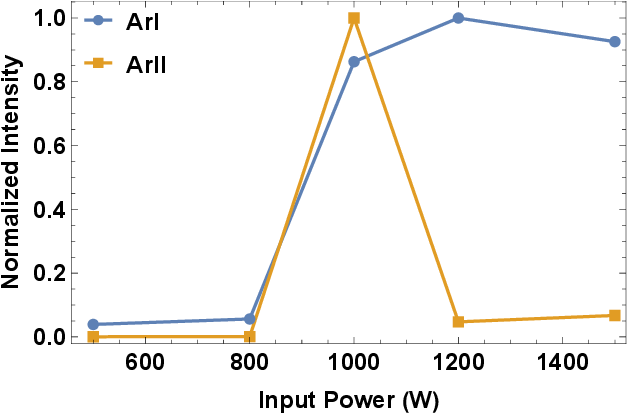}\\
(b)\\
\includegraphics[width=0.5\textwidth,angle=0]{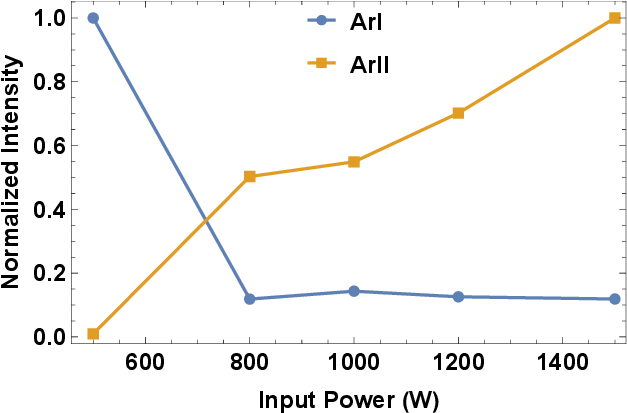}\\
(c)\\
\includegraphics[width=0.5\textwidth,angle=0]{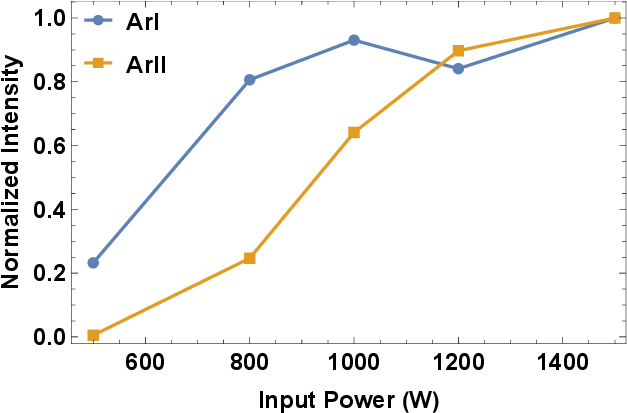}
\end{array}$
\end{center}
\caption{Normalized optical emission spectrometer (OES) intensities (choosing the maximum value from each curve and normalizing them for each figure of Fig.~\ref{fg_oes_power}) as function of input power: (a) $600$~G, (b) $800$~G, (c) $1000$~G.}
\label{fg_oes_power_axis}
\end{figure*}

We then explore the transition physics of blue-core helicon discharge via OES. Figure~\ref{fg_oes_power} shows the measured intensities for ArI (excited state of argon atom) and ArII (excited state of argon ion) and their dependences on the input power and magnetic field.
The wavelength corresponding to peak intensity of ArI is $\sim 763.2$~nm and for ArII it is $\sim 480.3$~nm. This wavelength shifts slightly when the input power and magnetic field vary but is negligible, i.e. the biggest shift is $0.2$~nm shown in Fig.~\ref{fg_oes_power}($b_1$). Every spectrum has a single peak, meaning that the emission corresponds to one single wavelength and is very clean (no mixture). To show more clearly the dependences of OES intensity on input power and magnetic field, we pick the maximum value from each curve and normalize the picked values for each figure. Results for ArI and ArII are also plotted together for comparison. As shown in Fig.~\ref{fg_oes_power_axis}, the dependence changes significantly for different magnetic field strengths.
For low field strength of $600$~G, the intensities of ArI and ArII both increase for power range of $500-1000$~W, however, for power higher than $1000$~W, the ArI intensity continues to grow while the ArII intensity starts to drop, which may be due to the impedance matching problem. Since the ArII intensity represents ion density and ArI intensity labels atom density, this implies that for the employed magnetic field strength of $600$~G, the discharge does not enter blue-core mode because for the power higher than $1000$~W ionization level drops down to give low density of ions but high density of neutral particles. For intermediate field strength of $800$~G, the ArII intensity grows with input power, whereas the ArI intensity drops continuously. For high field strength of $1000$~G, both the ArI intensity and ArII intensity increase for the full range of employed power. Overall, this indicates that high field strength is beneficial for blue-core discharge, which will be confirmed further as following. 

\begin{figure*}[ht]
\begin{center}
\includegraphics[width=1\textwidth,angle=0]{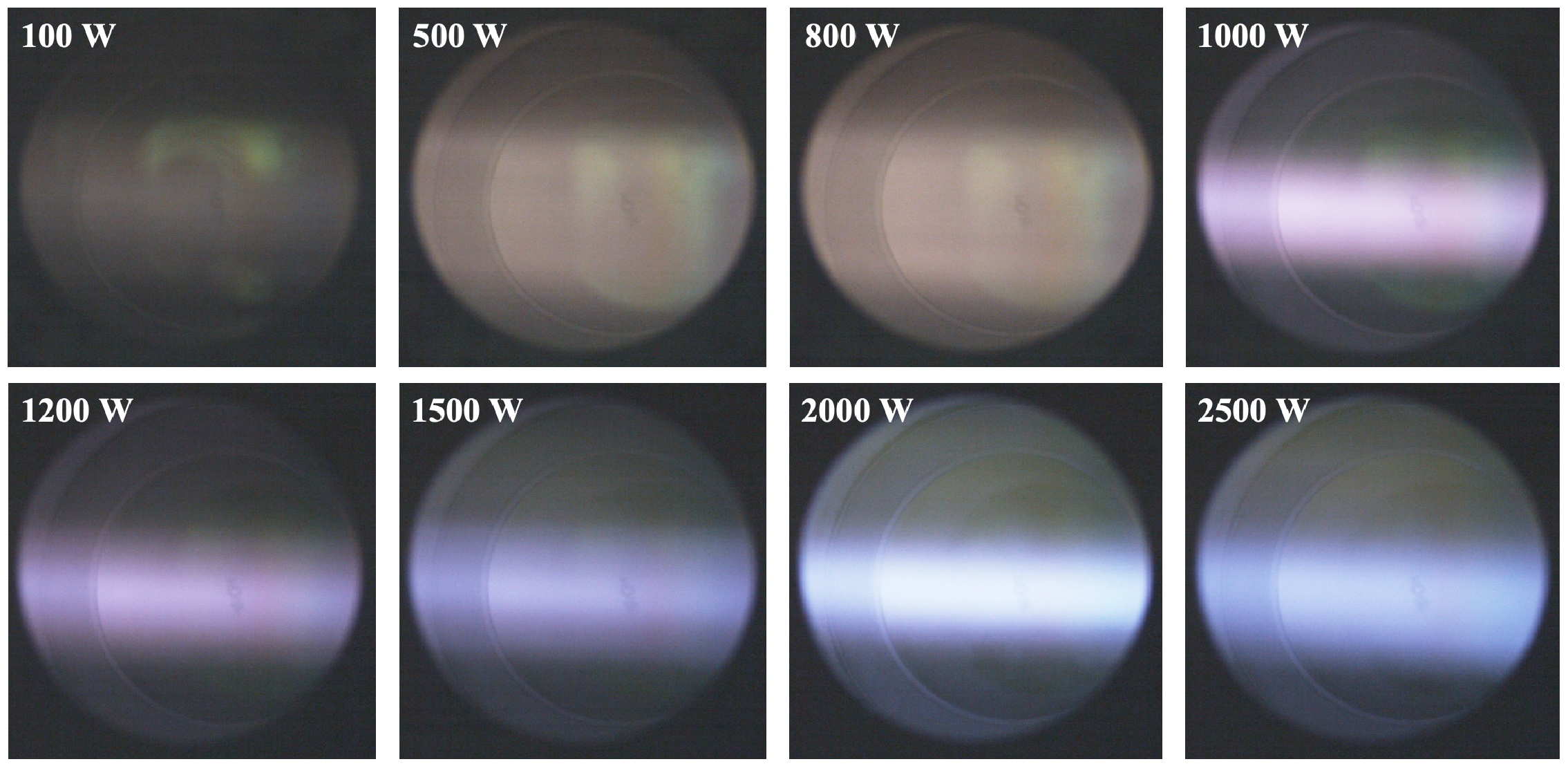}
\end{center}
\caption{Side-view images ($z=0.85$~m) of helicon discharge for different power levels of $100-2500$~W, magnetic field of $B_0=1000$~G, and background pressure of $0.19$~Pa.}
\label{fg_image_power}
\end{figure*}

\begin{figure*}[ht]
\begin{center}
\includegraphics[width=1\textwidth,angle=0]{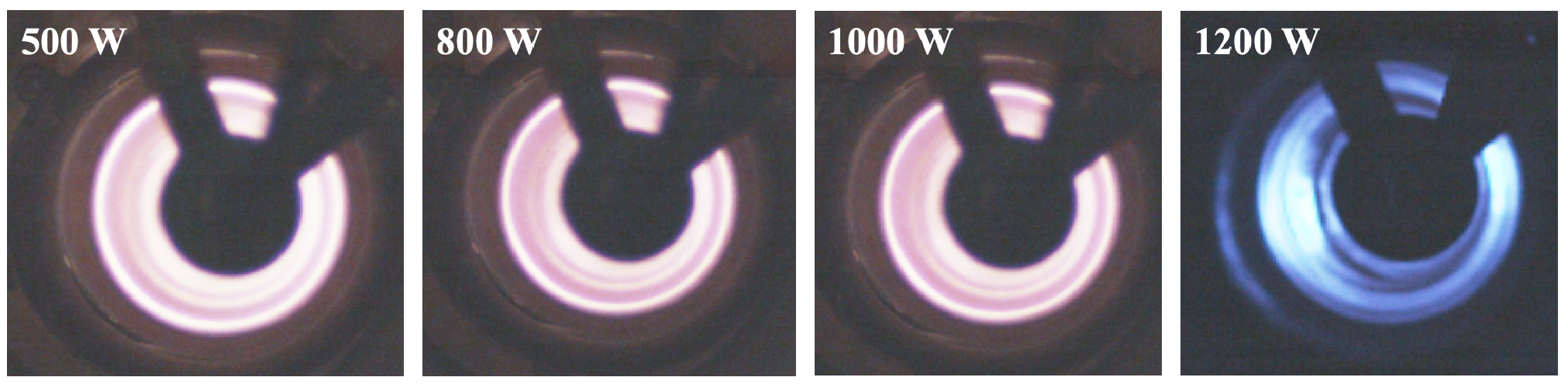}
\end{center}
\caption{End-view images ($z=0$~m) of helicon discharge for different power levels of $500-1200$~W, magnetic field of $B_0=1000$~G, and background pressure of $0.19$~Pa.}
\label{fg_image_end}
\end{figure*}

We also took standard high-speed videos of helicon discharge to capture the transitional details. Fig.~\ref{fg_image_power} shows the typical side-view images for confining magnetic field of $1000$~G. It can be seen that the discharge color changes from purple to blue as the input power increases, especially from $1000$~W to $2000$~W. For low field (e.g. $600$~G), however, the discharge color is purple all the time even for $2500$~W. Therefore, blue-core helicon discharge occurs easily for high field strength. These side-view images are taken at $z=0.85$~m, which is the only observation window that available in our experiments, although it is far away from the driving antenna. 
To show more details near the antenna, we took end-view images at $z=0$~m to capture the cross-sectional evolutions as shown by Fig.~\ref{fg_image_end}. The observed emission color in these end-view images corresponds to the entire plasma column and is captured in the line of sight. Additionally, the central region of the images is obscured by a dark object which is gas inlet valve and cannot be removed. Fig.~\ref{fg_image_end} indicates that the source plasma is already and completely blue for $1200$~W. This difference (it is still purple for $1200$~W in Fig.~\ref{fg_image_power}) can be attributed to the stronger ionization effect under the driving antenna. Overall, results imply that high field strength is beneficial for blue-core formation. More importantly, we observed from videos (available from the metadata repository of this paper) that the plasma column oscillates radially (side-view, yielding radial asymmetry) and experiences azimuthal instabilities (end-view) with a very high speed once entered the blue-core mode. These observations agree with other experiments which show that multiple instabilities may cause unstable discharge and form the blue-core helicon discharge\cite{Thakur:2014aa, Thakur:2015aa}. 

\subsection{Effects of Magnetic Field}
\begin{figure*}[ht]
\begin{center}$
\begin{array}{ll}
(a)&(b)\\
\includegraphics[width=0.495\textwidth,angle=0]{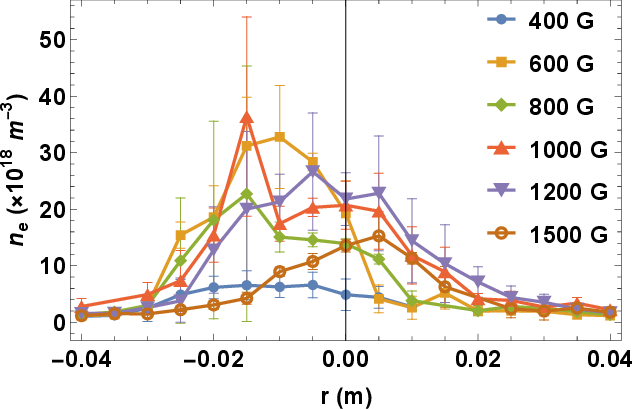}&\includegraphics[width=0.48\textwidth,angle=0]{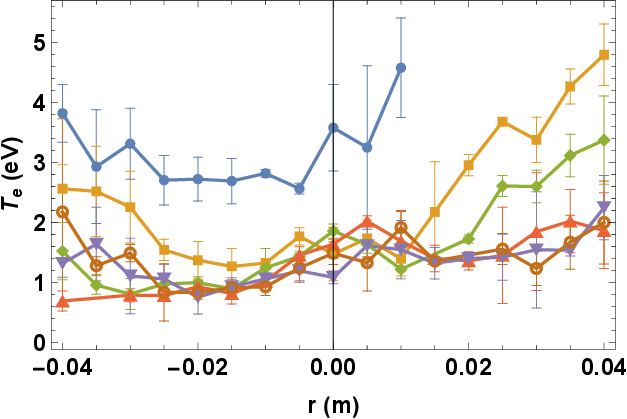}\\
\end{array}$
\end{center}
\caption{Radial profiles of helicon plasma for different external magnetic field strengths: (a) electron density, (b) electron temperature.}
\label{fg_mf}
\end{figure*}

\begin{figure*}[ht]
\begin{center}$
\begin{array}{ll}
(a)&(b)\\
\includegraphics[width=0.495\textwidth,angle=0]{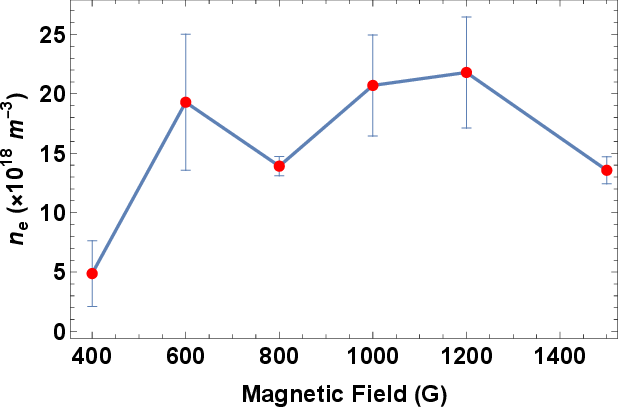}&\includegraphics[width=0.48\textwidth,angle=0]{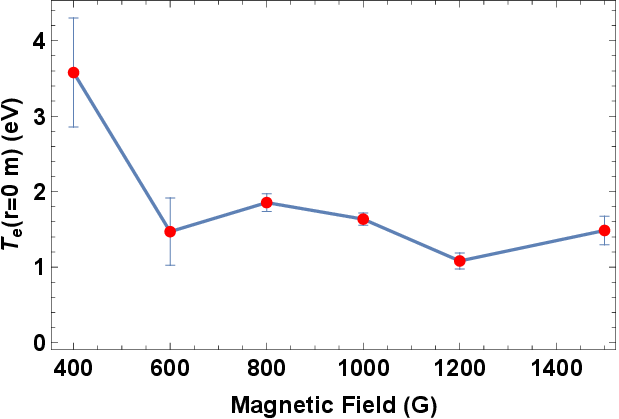}\\
\end{array}$
\end{center}
\caption{On-axis helicon plasma for different external magnetic field strengths: (a) electron density, (b) electron temperature.}
\label{fg_mf_axis}
\end{figure*}

\begin{figure*}[ht]
\begin{center}$
\begin{array}{ll}
(a)&(b)\\
\includegraphics[width=0.5\textwidth,angle=0]{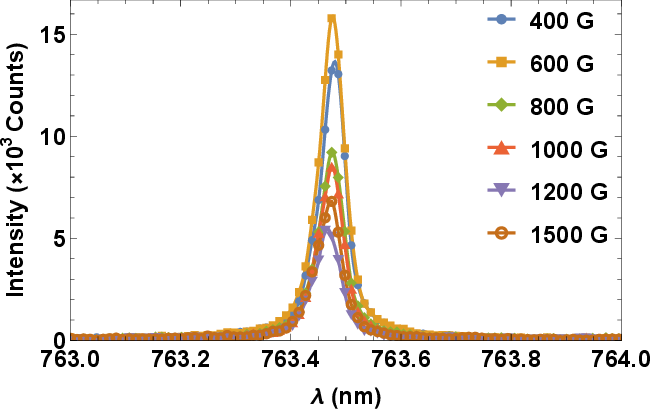}&\includegraphics[width=0.47\textwidth,angle=0]{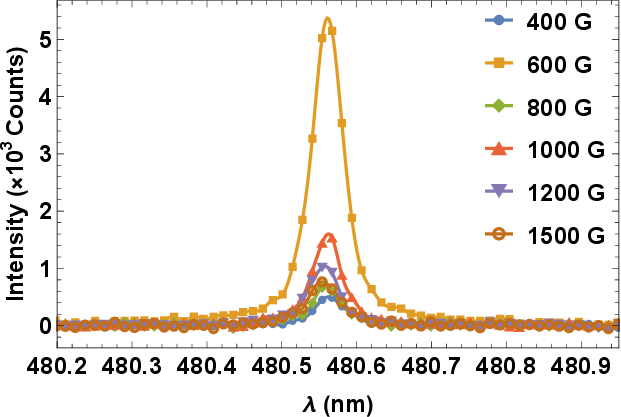}
\end{array}$
\end{center}
\caption{Measured optical emission spectrometer (OES) intensities of ArI (a) and ArII (b) for different magnetic field strengths.}
\label{fg_oes_field}
\end{figure*}

\begin{figure*}[ht]
\begin{center}
\includegraphics[width=0.5\textwidth,angle=0]{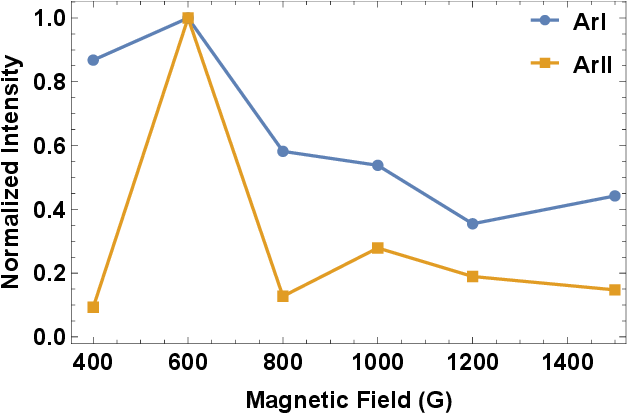}
\end{center}
\caption{Normalized optical emission spectrometer (OES) intensities as function of magnetic field strength.}
\label{fg_oes_field_norm}
\end{figure*}

\begin{figure*}[ht]
\begin{center}
\includegraphics[width=1\textwidth,angle=0]{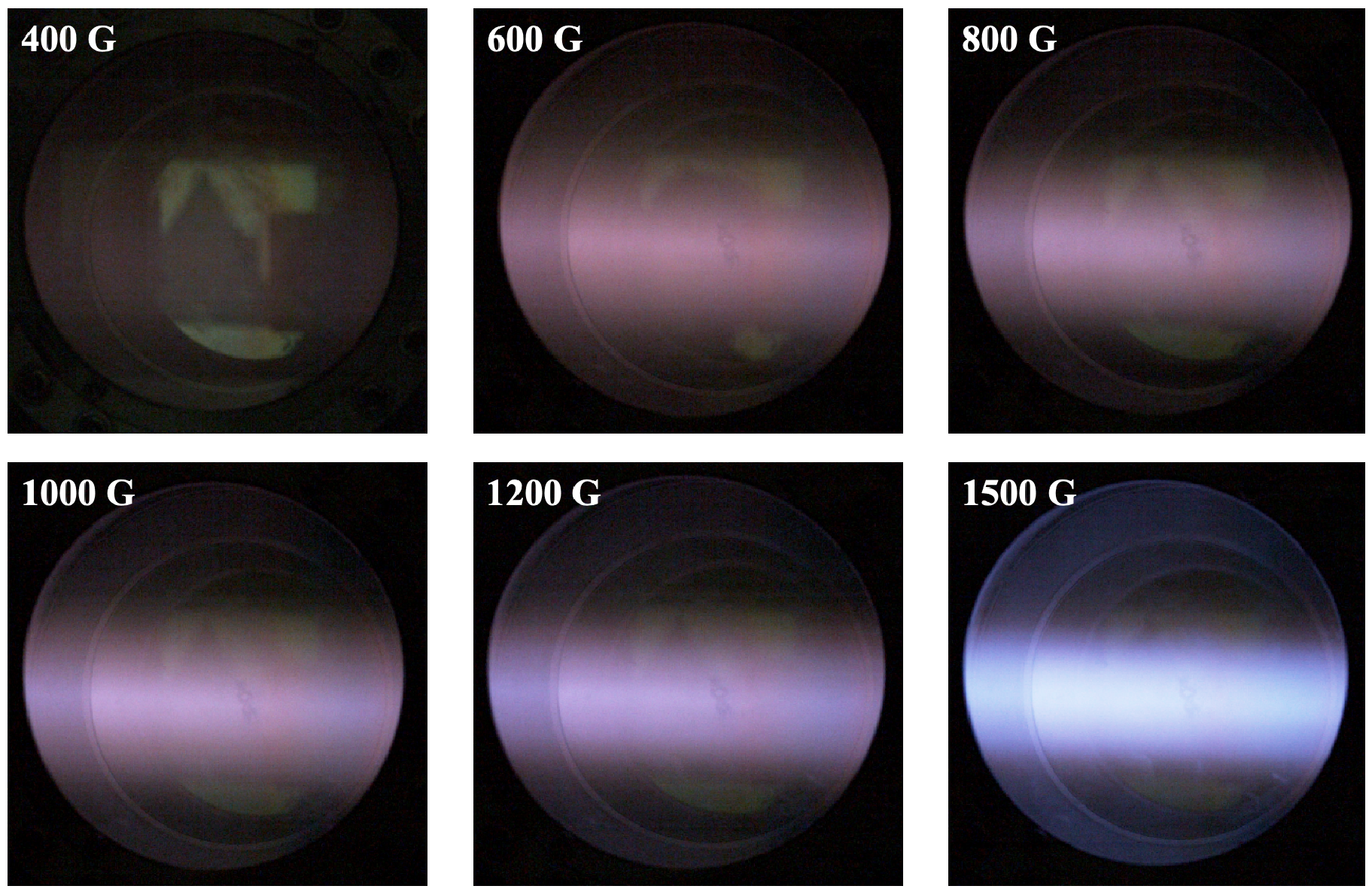}
\end{center}
\caption{Side-view images ($z=0.85$~m) of helicon discharge for different magnetic field of $400-1500$~G, input power of $1000$~W, and background pressure of $0.19$~Pa.}
\label{fg_image_field}
\end{figure*}

Then, we study the effects of confining magnetic field on the blue-core helicon discharge, i.e. ranging from $400$~G to $1500$~G.  The flow rate is also fixed to $60$~sccm as above, i.e. background pressure of $0.19$~Pa, and the power is set to be $1000$~W. Figure~\ref{fg_mf} shows the dependences of the radial profiles of electron density and temperature on the magnetic field strength. 
One can see that the overall magnitude of electron density increases with field strength, whereas for electron temperature this trend is opposite. Again, off-axis peak of electron density appears, especially for high magnetic field. To show these dependences more clearly, we choose the on-axis values and plot them in Fig.~\ref{fg_mf_axis}.
The driven mechanism of these opposite trends could be that strong magnetic field promotes blue-core mode discharge that enhances the ionization level and confines most plasma near axis to yield high electron density; however, better confinement also limits the radial diffusion which lowers the electron density near edge, so that the heating effects from external antenna into plasma column become weak to yield low electron temperature\cite{Chang2022}. The OES intensities of ArI and ArII and their dependences on magnetic field strength are shown in Fig.~\ref{fg_oes_field}. Again, every spectrum has single peak and the shift among peak locations for different field strengths is negligible. 
Figure~\ref{fg_oes_field_norm} shows more clearly the dependence of OES intensity on magnetic field strength through normalization of peaked values (via the same method as for Fig.~\ref{fg_oes_power_axis}).
We can observe that for both ArI and ArII their OES intensities grow first and then drop continuously. Figure~\ref{fg_image_field} displays the typical images of helicon discharge for magnetic field of $400-1500$~G, input power of $1000$~W, and background pressure of $0.19$~Pa. One can see the evolutionary formation of blue-core column, especially from $1000$~G to $1500$~G. The radial shaking of plasma column can be also observed from the video (available from the metadata repository of this paper), implying the existence of instabilities that causing unstable discharge. 

\subsection{Effects of Background Pressure}
\begin{figure*}[ht]
\begin{center}$
\begin{array}{ll}
(a)&(b)\\
\includegraphics[width=0.495\textwidth,angle=0]{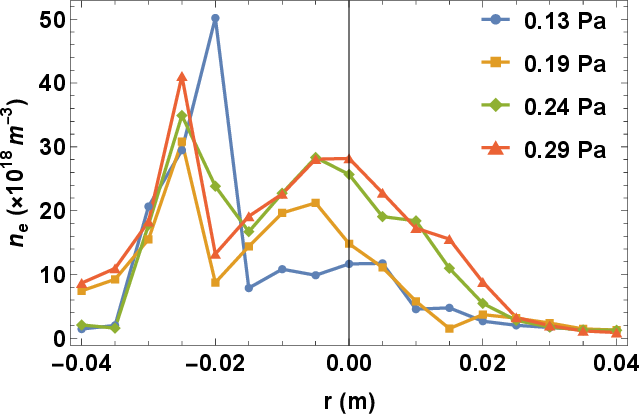}&\includegraphics[width=0.48\textwidth,angle=0]{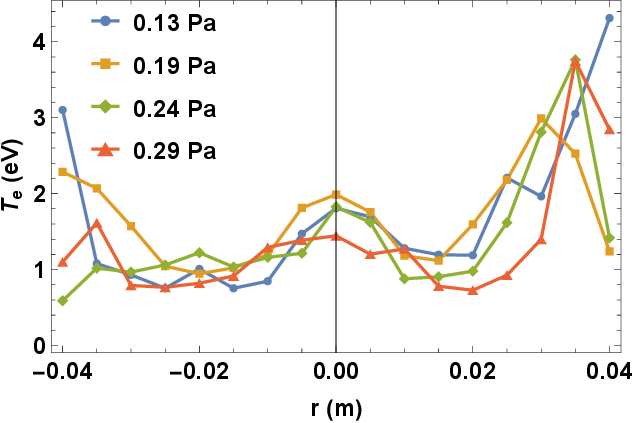}\\
\end{array}$
\end{center}
\caption{Dependence of radial plasma profiles on background pressure in terms of: (a) electron density, (b) electron temperature.}
\label{fg_trans_fl}
\end{figure*}

\begin{figure*}[ht]
\begin{center}$
\begin{array}{ll}
(a)&(b)\\
\includegraphics[width=0.485\textwidth,angle=0]{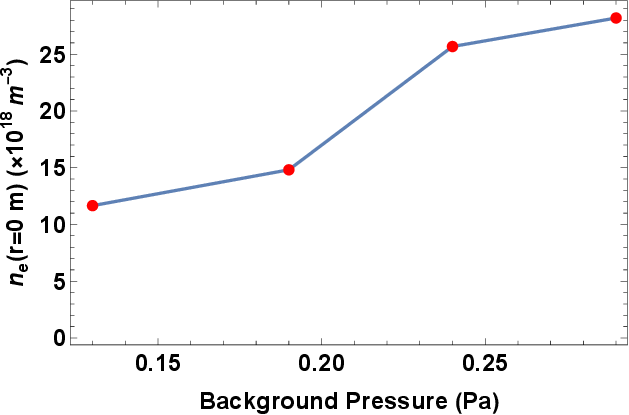}&\includegraphics[width=0.485\textwidth,angle=0]{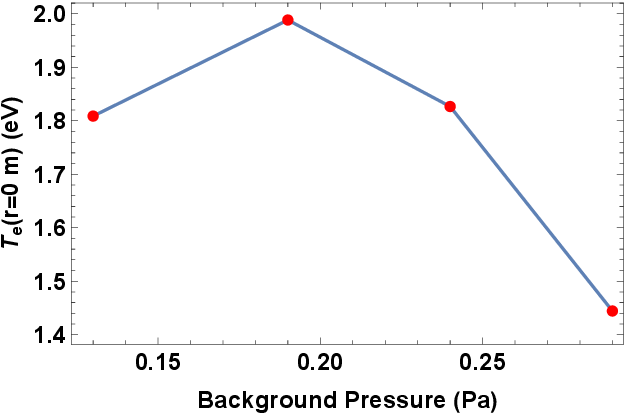}\\
\end{array}$
\end{center}
\caption{Variation of on-axis helicon plasma with background pressure in terms of: (a) electron density, (b) electron temperature.}
\label{fg_axis_fl}
\end{figure*}

\begin{figure*}[ht]
\begin{center}$
\begin{array}{ll}
(a)&(b)\\
\includegraphics[width=0.5\textwidth,angle=0]{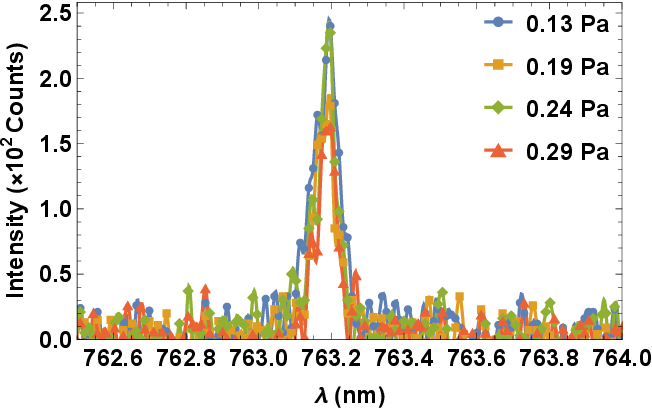}&\includegraphics[width=0.47\textwidth,angle=0]{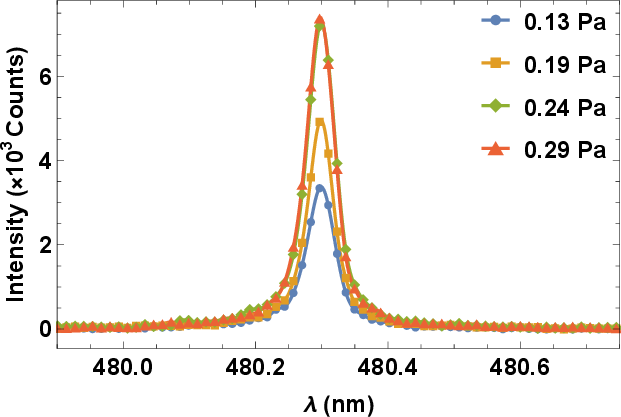}
\end{array}$
\end{center}
\caption{Measured optical emission spectrometer (OES) intensities of ArI (a) and ArII (b) for different background pressure levels.}
\label{fg_oes_flow}
\end{figure*}

\begin{figure*}[ht]
\begin{center}
\includegraphics[width=0.5\textwidth,angle=0]{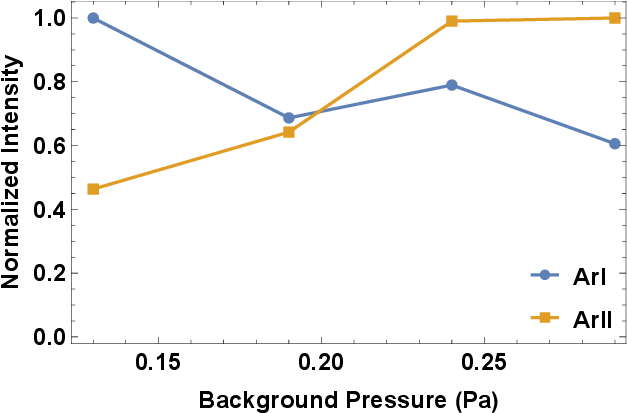}
\end{center}
\caption{Normalized optical emission spectrometer (OES) intensities as function of background pressure.}
\label{fg_oes_flow_norm}
\end{figure*}

As helicon discharge involves partially ionized plasmas, neutral particle dissipation and gas diffusion in principle effect the blue-core mode significantly. Thus, we finally study the effects of background pressure on the blue-core mode transition. Four typical pressure levels are considered: $0.13$~Pa, $0.19$~Pa, $0.24$~Pa, and $0.29$~Pa. Here, the input RF power and magnetic field are set to be $1000$~W and $1000$~G, respectively. Figure~\ref{fg_trans_fl} shows the radial profiles of electron density and temperature and Fig.~\ref{fg_axis_fl} presents their on-axis values, as the functions of background pressure.
We can see that the electron density increases with background pressure for the chosen range, while the electron temperature peaks around $0.19$~Pa, indicating the existence of best match among input power, magnetic field and background pressure for helicon discharge. Interestingly, the radial profile of electron temperature looks like a ``W" shape, i.e. minimizing around the edge of blue-core plasma column ($\pm 0.02$~m). 
Figure~\ref{fg_oes_flow} shows the OES intensities of ArI and ArII and their dependences on the background pressure.
These dependences are shown more clearly after normalization, as shown in Fig.~\ref{fg_oes_flow_norm}. One can draw conclusion that the variation of ArI intensity which decreases with background pressure is opposite to that of ArII intensity. It means that the fixed input power of $1000$~W and magnetic field strength of $1000$~G are high enough for the blue-core transition, so that most neutral particles can be ionized for the full range of background pressure considered here. 

\section{Numerical Simulation}
The wave field and power absorption of helicon discharge, which are not yet measured in the MPS-LD experiment, are important to understanding the transitional physics from non-blue-core mode to blue-core mode. To fill this gap, we shall compute them via an electromagnetic solver (EMS) which is verified by many experiments\cite{Chen:2006aa, Zhang:2008aa, Lee:2011aa, Chang:2012aa, Zhang:2023aa}.

\subsection{Governing Equations}
While details of EMS could be found in \cite{Chen:2006aa}, governing equations are given below. The EMS is based on two Maxwell's equations: Faraday's law and Ampere's law,
\small
\begin{equation}
\nabla\times\mathbf{E}=-\frac{\partial\mathbf{B}}{\partial t},
\label{eq14}
\end{equation}
\begin{equation}
\nabla\times\mathbf{B}=\mu_0\left(\mathbf{j}_a+\frac{\partial\mathbf{D}}{\partial t}\right),
\label{eq15}
\end{equation}
\normalsize
with $\mathbf{E}$ and $\mathbf{B}$ the wave electric and magnetic fields, respectively. The symbols of $\mu_0$ and $t$ are standard permeability of vacuum and time. The system is driven by the current density $\mathbf{j}_a$ of external antenna. Perturbations vary in form of $\exp[i(kz+m\theta-\omega t)]$, with $k$ the axial wave number, $m$ the azimuthal mode number and $\omega$ the driving frequency, for a right-hand cylindrical coordinate system $(r;\theta;z)$. The displacement vector $\mathbf{D}$ is linked to $\mathbf{E}$ via a cold-plasma dielectric tensor\cite{Ginzburg:1970aa},
\small
\begin{equation}
\mathbf{D}=\varepsilon_0[\varepsilon\mathbf{E}+i g(\mathbf{E}\times\mathbf{b})+(\eta-\varepsilon)(\mathbf{E}\cdot\mathbf{b})\mathbf{b}]. 
\label{eq16}
\end{equation}
\normalsize
Here, $\varepsilon_0$ is the permittivity of vacuum and $\mathbf{b}$ is the unit vector of external magnetic field ($\mathbf{b}=\mathbf{B_0}/B_0$). The dielectric tensor comprises three components: 
\small
\begin{equation}
\varepsilon=1-\sum_\alpha\frac{\omega+i\nu_\alpha}{\omega}\frac{\omega_{p\alpha}^2}{(\omega+i\nu_\alpha)^2-\omega_{c\alpha}^2},
\label{eq17}
\end{equation}
\begin{equation}
g=-\sum_\alpha\frac{\omega_{c\alpha}}{\omega}\frac{\omega_{p\alpha}^2}{(\omega+i\nu_\alpha)^2-\omega_{c\alpha}^2},
\label{eq18}
\end{equation}
\begin{equation}
\eta=1-\sum_\alpha\frac{\omega_{p\alpha}^2}{\omega(\omega+i\nu_\alpha)}.
\label{eq19}
\end{equation}
\normalsize
The subscript $\alpha$ labels the species of particles, i. e. ion and electron, and the plasma frequency $\omega_{p\alpha}=\sqrt{n_\alpha q_\alpha^2/\varepsilon_0 m_\alpha}$ and cyclotron frequency $\omega_{c\alpha}=q_\alpha B_0/m_\alpha$ are standard definitions. The phenomenological collision frequency $\nu_\alpha$ accounts for collisions between electrons, ions and neutrals, where background pressure is implemented. For the half-turn helical antenna considered in MPS-LD experiment, $\mathbf{j}_a$ has three components: 
\small
\begin{equation}
\begin{array}{ll}
j_{ar}=0, 
\label{eq20}
\end{array}
\end{equation}
\begin{equation}
\begin{array}{ll}
j_{a\theta}=&I_a\frac{e^{im\pi}-1}{2}\delta(r-R_a)\left\{\frac{i}{m\pi}\left[\delta(z-z_a)+\delta(z-z_a-L_a)\right]\right.\\
\\
&\left.+\frac{H(z-z_a)H(z_a+L_a-z)}{L_a}e^{-im\pi[1-(z-z_a)/L_a]}\right\},
\label{eq21}
\end{array}
\end{equation}
\begin{equation}
\begin{array}{ll}
j_{az}=I_a\frac{e^{-im\pi[1-(z-z_a)/L_a]}}{\pi R_a}\frac{1-e^{im\pi}}{2}\delta(r-R_a)\times H(z-z_a)H(z_a+L_a-z). 
\label{eq22}
\end{array}
\end{equation}
\normalsize
Here, the subscript $a$ denotes the antenna, i. e. $L_a$ the length, $R_a$ the radius, $z_a$ the distance to left endplate, $I_a$ the magnitude of antenna current, and $H$ is the Heaviside step function. The boundary conditions enclosing the model are formed by assuming that the tangential components of $\mathbf{E}$ vanish on the surface of chamber walls:
\small
\begin{equation}
E_\theta(R, z)=E_z(R, z)=0, 
\label{eq23}
\end{equation}
\begin{equation}
E_r(r, 0)=E_\theta(r, 0)=0, 
\label{eq24}sd
\end{equation}
\begin{equation}
E_r(r, L)=E_\theta(r, L)=0,
\label{eq25}
\end{equation}
\normalsize
where $R$ and $L$ are the radius and length of chamber, respectively. The aforementioned equations are solved numerically via a finite difference method based on four staggered rectangular grids\cite{Chen:2006aa} and the following input parameters and conditions.

\subsection{Input Parameters}
\begin{figure*}[ht]
\begin{center}
\includegraphics[width=\textwidth,angle=0]{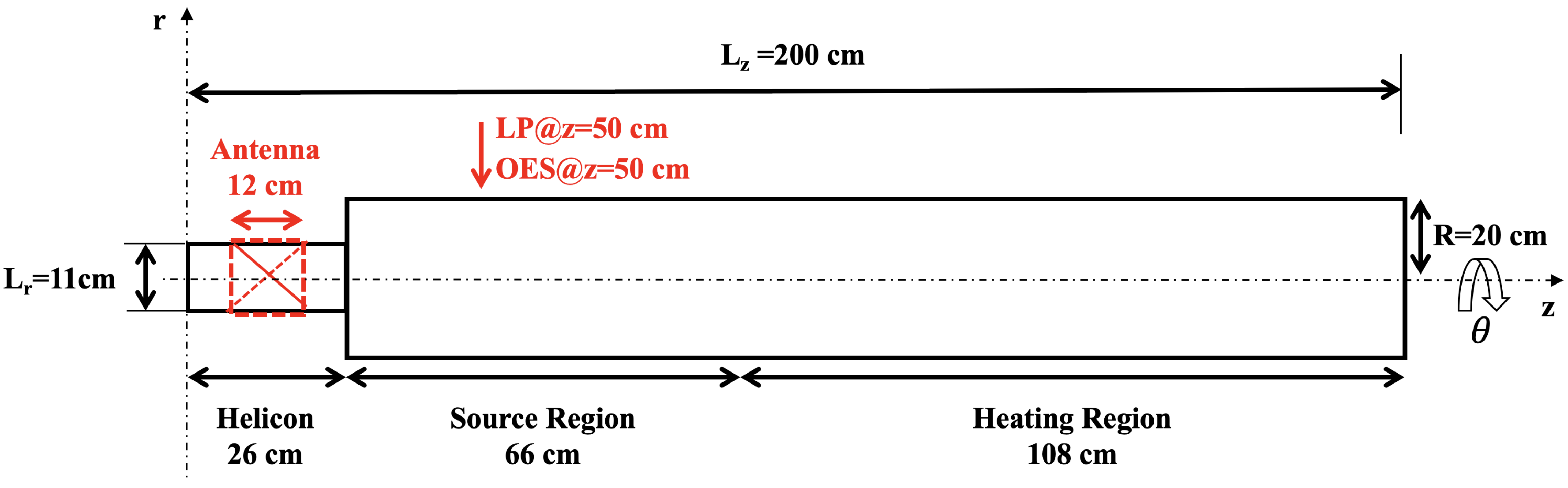}
\end{center}
\caption{Schematic of computational domain for EMS (ElectroMagnetic Solver).}
\label{fg_ems_domain}
\end{figure*}

\begin{figure*}[ht]
\begin{center}$
\begin{array}{lll}
(a_1)&(b_1)&(c_1)\\
\hspace{-0.2cm}\includegraphics[width=0.31\textwidth,angle=0]{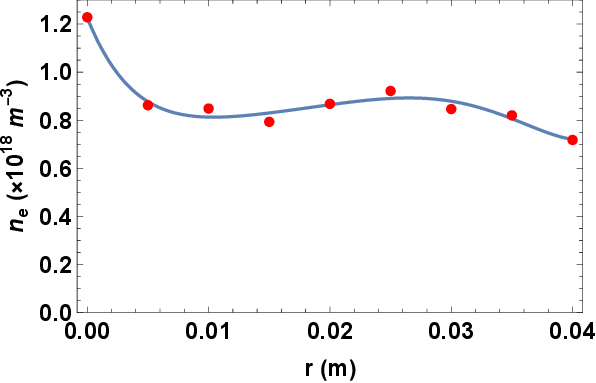}&\hspace{-0.2cm}\includegraphics[width=0.31\textwidth,angle=0]{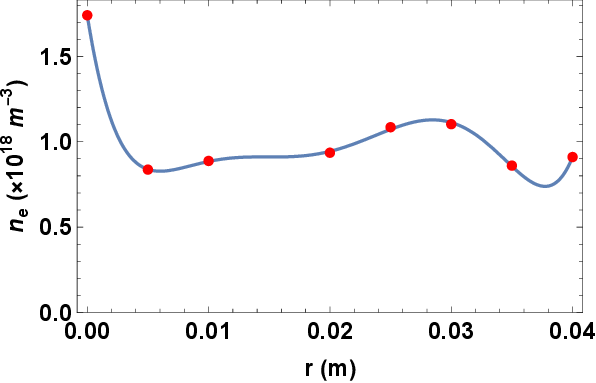}&\hspace{-0.2cm}\includegraphics[width=0.31\textwidth,angle=0]{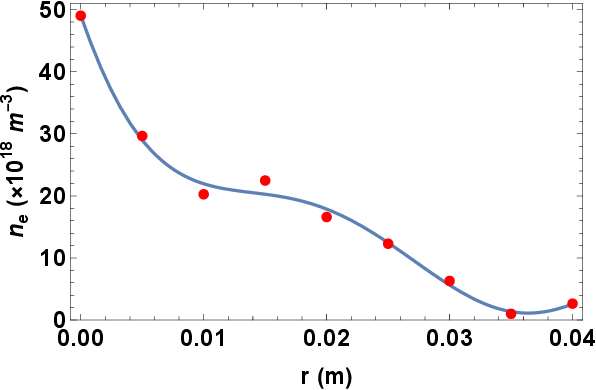}\\
(a_2)&(b_2)&(c_2)\\
\includegraphics[width=0.3\textwidth,angle=0]{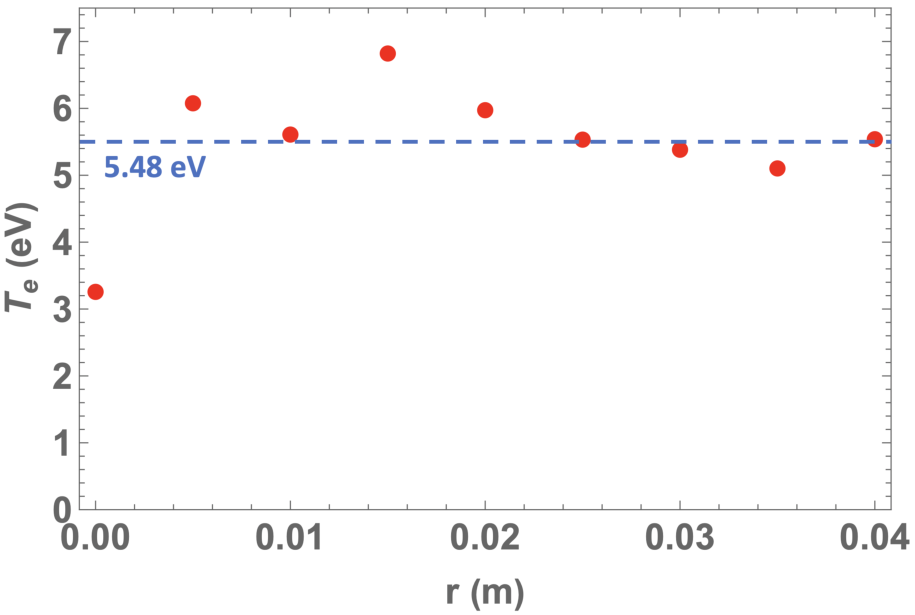}&\includegraphics[width=0.3\textwidth,angle=0]{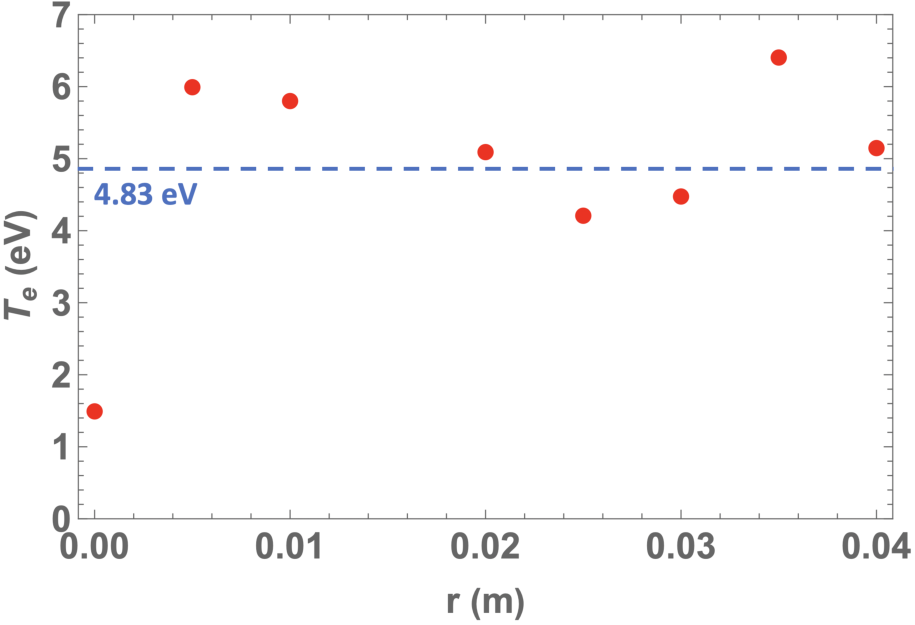}&\hspace{-0.2cm}\includegraphics[width=0.31\textwidth,angle=0]{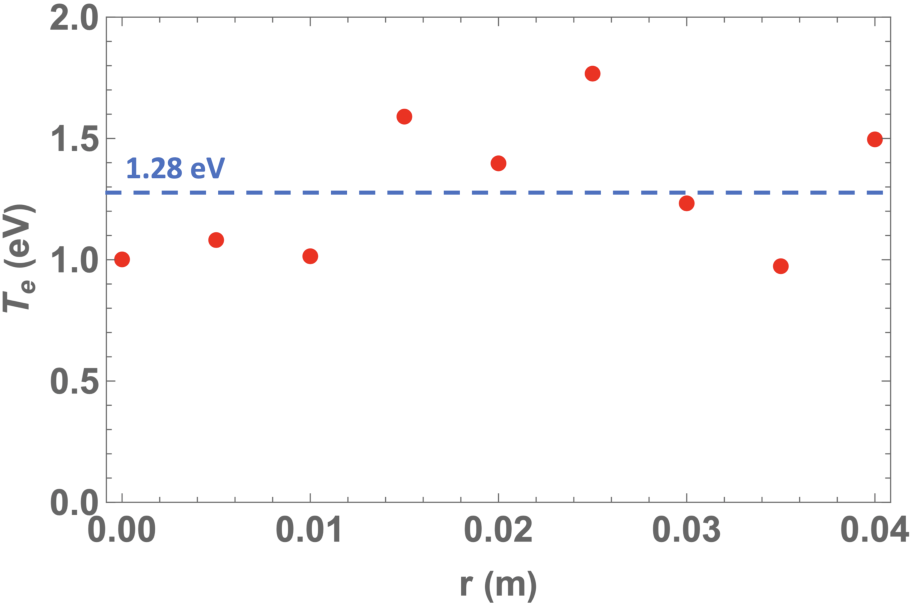}
\end{array}$
\end{center}
\caption{Radial profiles of plasma density (upper) and temperature (lower) for three power levels: (a) $890$~W, (b) $930$~W, (c) $990$~W, employed for the EMS computations (red points are experimental data and fitted lines (solid \& dashed) are inputs for EMS).}
\label{fg_ems_input}
\end{figure*}

Figure~\ref{fg_ems_domain} displays the computational domain employed by the EMS code. For the interest of present work, we choose the first $2$~m of MPS-LD, i.e. $z=0-2$~m in Fig.~\ref{fg_mps-ld}, which is immersed in a uniform magnetic field of the strength $B_0=0.1$~T. The working gas is argon as used in the experiment. Particularly, we choose three typical cases: $890$~W before blue-core transition, $930$~W during blue-core transition, and $990$~W after blue-core transition, referring to Fig.~\ref{fg_trans_com}, Fig.~\ref{fg_map_ne} and Fig.~\ref{fg_map_te}. The antenna current is set accordingly as well. 
The radial profiles of electron density and temperature for these three power levels are shown in Fig.~\ref{fg_ems_input} and input directly to the EMS code, i.e. fitted lines (solid and dashed) with the experimental data in Fig.~\ref{fg_map_ne}(a) and Fig.~\ref{fg_map_te}(a). Here, the radial asymmetry which is likely caused by unstable discharge is removed, through locating the peak density onto axis. 
The electron temperature is averaged through the whole radial domain to give a constant value, because non-uniform temperature is not supported yet by the current version of the EMS code. Moreover, due to the limit of axial diagnostics, full data of electron density and temperature in the axial direction are not available. However, experimental observations (e.g. Fig.~\ref{fg_image_power} and Fig.~\ref{fg_image_field}) indicate good axial uniformity which is thus considered in the EMS computations. 

\subsection{Computed Results}
\begin{figure*}[ht]
\begin{center}$
\begin{array}{ll}
(a_1)&(b_1)\\
\includegraphics[width=0.48\textwidth,angle=0]{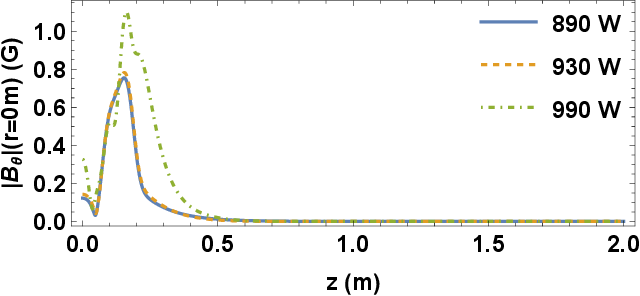}&\includegraphics[width=0.49\textwidth,angle=0]{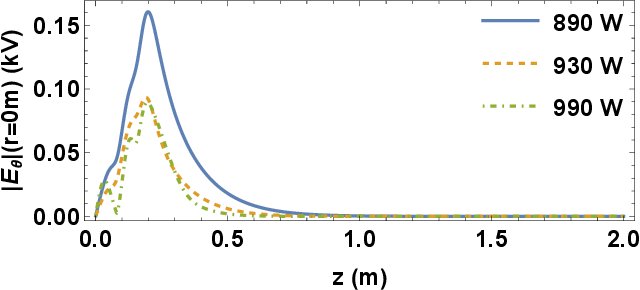}\\
(a_2)&(b_2)\\
\includegraphics[width=0.48\textwidth,angle=0]{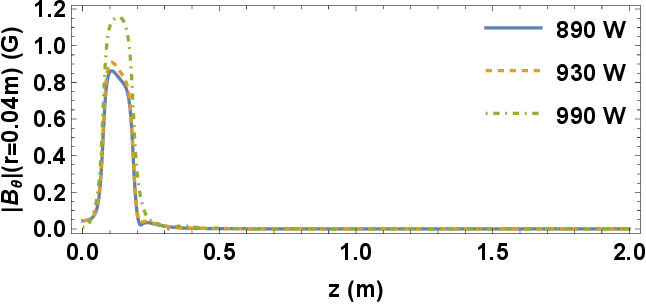}&\hspace{0.4cm}\includegraphics[width=0.465\textwidth,angle=0]{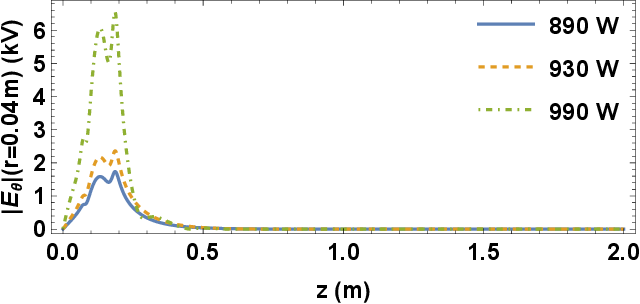}\\
\end{array}$
\end{center}
\caption{Axial profiles of wave magnetic field (a) and wave electric field (b) measured on axis (upper) and near edge (lower) from the EMS computations.}
\label{fg_ems_axial}
\end{figure*}

\begin{figure*}[ht]
\begin{center}$
\begin{array}{ll}
(a_1)&(b_1)\\
\includegraphics[width=0.475\textwidth,angle=0]{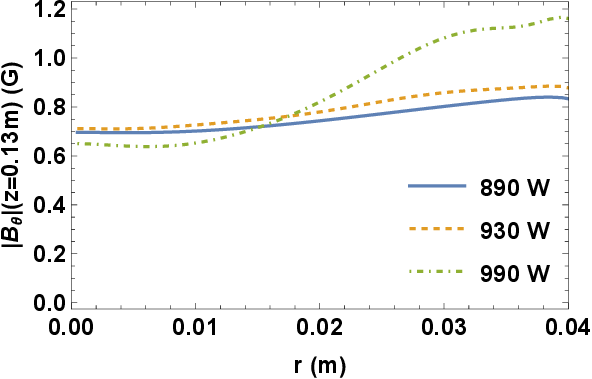}&\includegraphics[width=0.495\textwidth,angle=0]{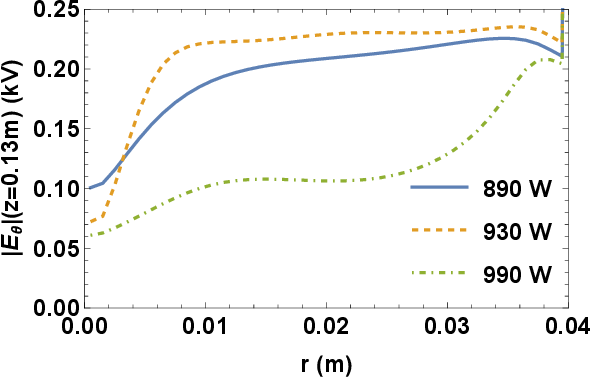}\\
(a_2)&(b_2)\\
\includegraphics[width=0.475\textwidth,angle=0]{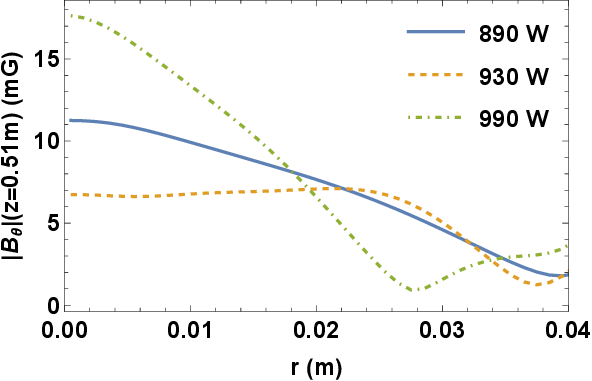}&\hspace{0.35cm}\includegraphics[width=0.475\textwidth,angle=0]{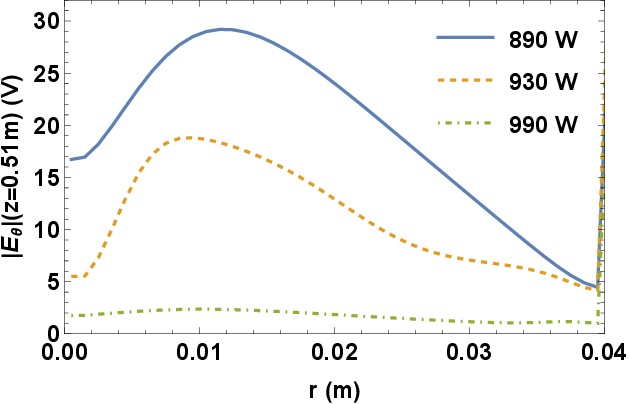}\\
\end{array}$
\end{center}
\caption{Radial profiles of wave magnetic field (a) and wave electric field (b) measured under the middle of antenna (upper, $z=0.13$~m) and at the location of Langmuir probe and OES (lower, $z=0.51$~m), computed by the EMS code.}
\label{fg_ems_radial}
\end{figure*}

\begin{figure*}[ht]
\begin{center}$
\begin{array}{lll}
(a_1)&(b_1)&(c_1)\\
\includegraphics[width=0.31\textwidth,angle=0]{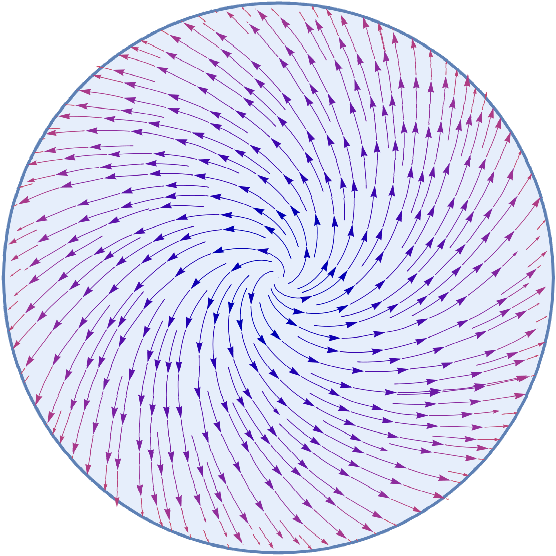}&\includegraphics[width=0.31\textwidth,angle=0]{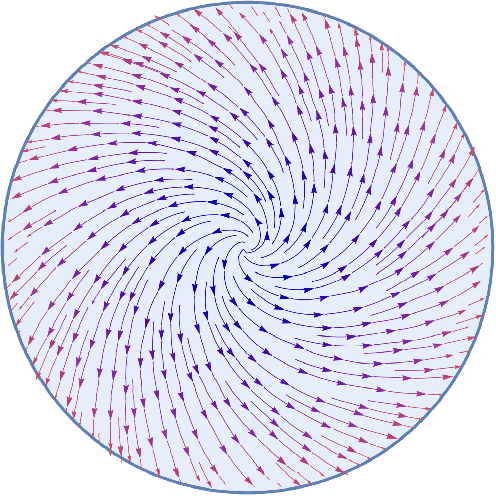}&\includegraphics[width=0.31\textwidth,angle=0]{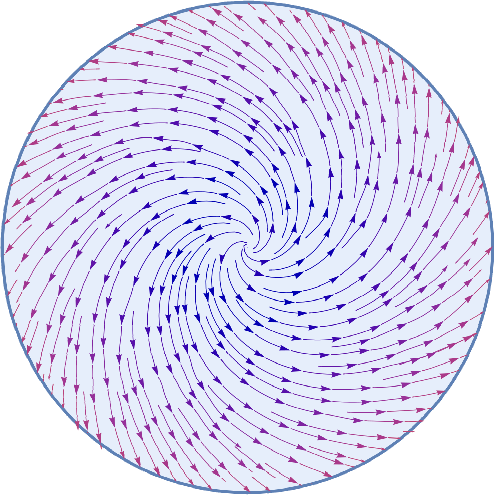}\\
(a_2)&(b_2)&(c_2)\\
\includegraphics[width=0.31\textwidth,angle=0]{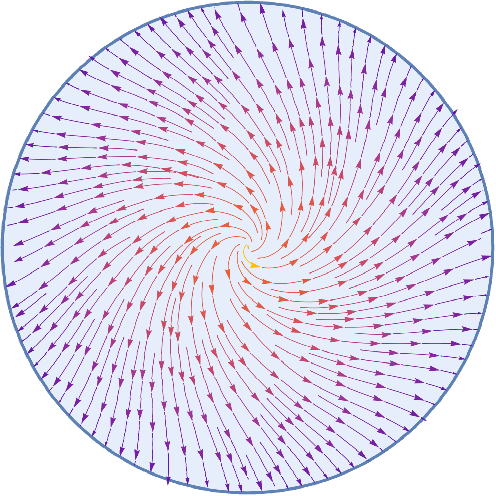}&\includegraphics[width=0.31\textwidth,angle=0]{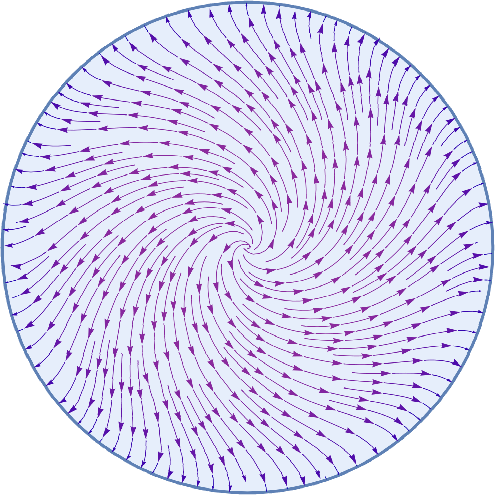}&\includegraphics[width=0.31\textwidth,angle=0]{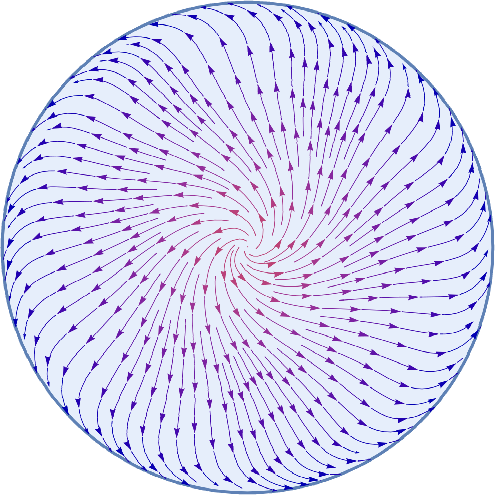}
\end{array}$
\end{center}
\caption{Stream plots of the cross-sectional wave field under the antenna (upper, $z=0.13$~m) and in the location of Langmuir probe/OES (lower, $z=0.51$~m) for three power levels: (a) $890$~W, (b) $930$~W, (c) $990$~W, computed from the EMS code.}
\label{fg_ems_vector}
\end{figure*}

\begin{figure*}[ht]
\begin{center}
\includegraphics[width=0.99\textwidth,angle=0]{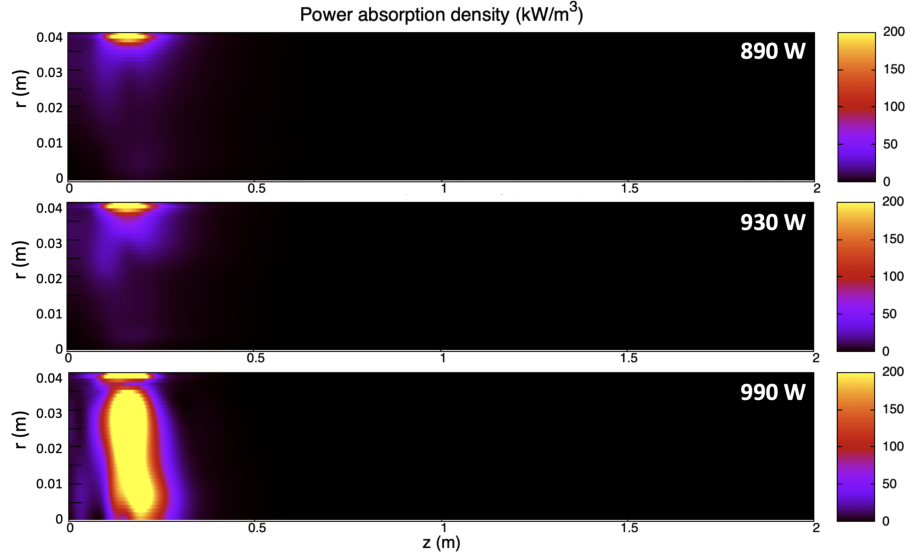}
\end{center}
\caption{2D power absorption density for three power levels: $890$~W, $930$~W, $990$~W, computed by the EMS code.}
\label{fg_ems_power}
\end{figure*}

Figure~\ref{fg_ems_axial} shows the computed axial profiles of wave magnetic field and wave electric field for two radial locations, i.e. on axis ($r=0$~m, inside core) and near edge ($r=0.04$~m, outside core), respectively. It can be seen that wave fields for $890$~W and $930$~W are similar but significantly different from that of $990$~W. Specifically, inside the core (on axis), wave magnetic field increases from non-blue-core mode ($890$~W and $930$~W) to blue-core mode ($990$~W), whereas the wave electric field drops during this transition; outside the core (near edge), both wave magnetic field and wave electric field increase during the blue-core transition. The different phenomena may be attributed to different density levels on axis and near edge. Further, different from previous study which shows that wave propagates better inside the core than that outside\cite{Chang:2022aa}, here the trend is not clear, i.e. the trend from wave magnetic field is opposite to that from wave electric field inside the core. 
The radial profiles of wave field are given in Fig.~\ref{fg_ems_radial}. It shows that the wave field structure changes significantly from non-blue-core mode ($890$~W and $930$~W) to blue-core mode ($990$~W). Specifically, under the driving antenna ($z=0.13$~m), the radial profile of wave field becomes more hollow, i.e. decreased near axis but increased near edge, when the blue-core transition occurs. However, where it is far away from the antenna ($z=0.51$~m in the probe location), the radial profiles of wave field becomes more peaked, i.e. increased near axis but decreased near edge, during the blue-core transition. This opposite trend at two axial locations may be related to other mode transitions or coupling effects during axial decay, and is not straightforward to explain. We thus leave it for future study due to the limited scope of present work. In terms of wave magnitude, again, the trend from wave magnetic field is overall opposite to that from wave electric field during the blue-core transition.
To show more clearly the cross-sectional structure of wave field, we take $B_r$ and $B_\theta$ for example and transform the cylindrical coordinate to Cartesian coordinate to produce stream plots, as shown in Fig.~\ref{fg_ems_vector}. Again, results for non-blue-core mode ($890$~W and $930$~W) are similar but different from that of blue-core mode ($990$~W), which is displayed more clearly in the downstream region (lower row of Fig.~\ref{fg_ems_vector}). Difference can be also observed between the upper row (location of antenna) and lower row (location of Langmuir probe and OES). 
Figure~\ref{fg_ems_power} presents 2D power absorption density for the three power levels. 
One can see that blue-core mode ($990$~W) has much more efficient power absorption from external antenna towards internal axis than non-blue-core mode ($890$~W and $930$~W). Overall, however, most power is absorbed under the antenna, showing axially decaying feature.

\section{Conclusion}
In this work, we explored the transitional features of blue-core helicon discharge through both experimental measurements on a recently built advanced helicon device (MPS-LD) and numerical computations via a well benchmarked full wave code (EMS). Details in terms of electron density, electron temperature, emission spectrum, optical videos, wave field structure, and power absorption density are presented. The effects of input RF power, external magnetic field, and background pressure on the blue-core transition are investigated as well. Key and novel findings are summarized below: 
\begin{itemize}
\item the jump direction of electron density (from low level to high level) is opposite to that of electron temperature (from high level to low level);
\item electron density increases significantly and the radial profile becomes localized near the axis when the blue-core transition occurs, with the peak density located off-axis;
\item the radial profile of electron temperature looks like a ``W" shape, i.e. minimizing around the edge of blue-core column;
\item electron density increases with background pressure, while electron temperature peaks around certain pressure level, indicating there exists a best match among input power, magnetic field and background pressure for blue-core helicon discharge;
\item high-speed videos show that the blue-core plasma column oscillates radially and experiences azimuthal instabilities with high rate once entered blue-core mode; 
\item wave field magnitude changes significantly from non-blue-core mode to blue-core mode, and its radial structure differs from antenna to downstream; 
\item the trend of power dependence of wave magnetic field is overall opposite to that of wave electric field during the blue-core mode transition; 
\item most power is absorbed under the antenna, which shows clear axial decay feature. 
\end{itemize}
These comprehensive explorations of blue-core helicon discharge show various interesting features and physics which are not presented before, and are believed to be of important value for the helicon community to understand the formation physics of blue-core helicon plasma and use it for practical purposes. Future research will be devoted to the energetic electron measurements using an upgraded retarding potential analyzer (RPA) and the instabilities and transport physics (including transport barrier) during the blue-core helicon plasma formation. Moreover, helicon discharges, particularly at high power, involve complex nonlinear phenomena. These extend beyond traditional plasma-wave interactions to include dynamic thermal dissipation and equilibrium challenges arising from high ionization degrees and elevated ion temperatures, which will be studied too.

\ack
This work is supported by National Natural Science Foundation of China (92271113, 12411540222, 12481540165, 12122503, and 12235002), Chongqing Natural Science Foundation (CSTB2025NSCQ-GPX0725), Fundamental Research Funds for Central Universities (2022CDJQY-003), Chongqing Entrepreneurship and Innovation Support Program for Overseas Returnees (CX2022004), Dalian Science and Technology Talents Program (2022RJ11), and Xingliao Talent Project (XLYC220).

\section*{Data Availability Statement}
The data that support the findings of this study are available from the corresponding authors upon reasonable request.

\section*{ORCID IDs}
\raggedright
Lei Chang: https://orcid.org/0000-0003-2400-1836

Shi-Jie Zhang: https://orcid.org/0009-0003-9565-6095

Jin-Tao Wu: https://orcid.org/0009-0001-0026-8491 

Yi-Wei Zhang: https://orcid.org/0009-0000-4049-4568

Chao Wang: https://orcid.org/0009-0004-9169-5975

Yao Peng: https://orcid.org/0009-0000-2371-2703

Shuai-Shuai Gao: https://orcid.org/0009-0009-1314-2706

Chang-Jiang Sun: https://orcid.org/0000-0002-2534-5473

Qi Wang: https://orcid.org/0000-0002-3455-3098

Chao-Feng Sang: https://orcid.org/0000-0002-6861-5242

Shogo Isayama: https://orcid.org/0000-0002-7531-8211

Shin Jae You: https://orcid.org/0000-0001-8005-7880

\section*{References}
\bibliographystyle{unsrt}

\end{document}